\documentclass[english]{article}

\usepackage[colorlinks,bookmarksopen,bookmarksnumbered,allcolors=red]{hyperref}
\usepackage{xcolor}
\usepackage{graphicx}
\usepackage{geometry}
\usepackage{amsmath, amssymb, bm}
\usepackage{authblk}
\usepackage{listings}
\renewcommand{\lstlistingname}{Listing}
\usepackage{placeins}
\definecolor{bluegreen}{RGB}{46,141,131}
\definecolor{darkgreen}{RGB}{46,139,87}
\definecolor{darkred}{RGB}{219,7,61}
\definecolor{darkblue}{RGB}{0,0,137}

\title{\textbf{Joint Modeling of Longitudinal Measurements and Time-to-event Outcomes Using BUGS}}

\author[,1]{Taban Baghfalaki\thanks{Corresponding author: taban.baghfalaki@u-bordeaux.fr}}
\author[2]{Mojtaba Ganjali}
\author[1]{Antoine Barbieri}
\author[1]{Hélène Jacqmin-Gadda}
\affil[1]{Inserm, Research Center U1219, Univ. Bordeaux, ISPED, F33076 Bordeaux, France}
\affil[2]{Department of Statistics, Faculty of Mathematical Sciences, Shahid Beheshti University, Tehran, Iran}

\date{}

\begin{document}

\maketitle

\begin{abstract}
The objective of this paper is to provide an introduction to the principles of Bayesian joint modeling of longitudinal measurements and time-to-event outcomes, as well as model implementation using the BUGS language syntax. This syntax can be executed directly using OpenBUGS or by utilizing convenient functions to invoke OpenBUGS and JAGS from R software. In this paper, all details of joint models are provided, ranging from simple to more advanced models. The presentation started with the joint modeling of a Gaussian longitudinal marker and time-to-event outcome. The implementation of the Bayesian paradigm of the model is reviewed. The strategies for simulating data from the JM are  also discussed. A proportional hazard model with various forms of baseline hazards, along with the discussion of all possible association structures between the two sub-models are taken into consideration.
The paper covers  joint models with multivariate longitudinal measurements,
zero-inflated longitudinal measurements, competing risks, and time-to-event with cure fraction. The models are illustrated by the analyses of several real data sets.
All simulated and real data and code are available at  \url{https://github.com/tbaghfalaki/JM-with-BUGS-and-JAGS}.
\end{abstract}

\section{Introduction}
\label{sec::intro}

Longitudinal and survival data analysis are among the most applied domains of statistics and biostatistics. The key issues for longitudinal data analysis are taking into account the within-subject correlation and handling missing data  \cite{diggle2002analysis,molenberghs2006longitudinal,fitzmaurice2008longitudinal}. On the other hand, survival analysis deals with time-to-event outcomes, which often involve incomplete data due to censoring  \cite{kleinbaum1996survival,klein2016handbook}. Since many medical and epidemiological studies record longitudinal measurements until a time-to-event outcome occurs, there has been a convergence of  these two different areas of statistics  toward a new field known as joint modeling of longitudinal measurements and survival analysis  \cite{rizopoulos2012joint,elashoff2016joint}. Therefore, various challenges in both areas, such as accounting for non-ignorable missingness in longitudinal data analysis or handling survival analysis with time dependent covariates, are tackled in the joint modeling. 

The popularity of joint modeling of longitudinal measurements and time-to-event outcomes has led to the development of many R packages for implementing  joint models. One such package is \texttt{JM}  \cite{JM}, which can fit joint models for a single continuous longitudinal outcome and a time-to-event outcome using maximum likelihood  approach. \texttt{joineR} implements the joint modeling approach proposed by Henderson et al.  \cite{henderson2000joint} and Williamson et al.  \cite{williamson2008joint}, which includes a longitudinal outcome as well as time-to-event or competing risk outcomes. The longitudinal outcome is modeled using a linear mixed-effects model, and a  proportional hazards regression model (with time-varying covariates) is considered for modeling the event time. The association is captured by a latent Gaussian process, and the model is estimated using the EM algorithm.
\texttt{joineRML} implements an extension of the joint modeling approach, proposed by Henderson et al.  \cite{henderson2000joint}, for analyzing multiple Gaussian longitudinal measures. In this package, the multiple longitudinal outcomes are modeled using a multivariate version of the Laird and Ware linear mixed model. \texttt{JMBayes}  \cite{rizopoulos2020package} implements joint modeling for single or multiple longitudinal outcomes and time-to-event data using MCMC. Another package, \texttt{JMBayes2}  \cite{rizopoulos2022jmbayes2} is a comprehensive tool for implementing Bayesian joint models for longitudinal and time-to-event data. It can accommodate multiple longitudinal outcomes of different types. Additionally, it covers competing risks, multi-state, and recurrent event processes for the time-to-event model. \texttt{JSM}  \cite{xu2020semi} performs maximum likelihood estimation (MLE) for the semi-parametric joint modeling of survival and longitudinal data. It is based on the non-parametric multiplicative random effects model proposed by Ding and Wang  \cite{ding2008modeling}. On the other hand, \texttt{INLAjoint}  \cite{rustand2022denisrustand} fits joint models for multivariate longitudinal markers and survival outcomes using Integrated Nested Laplace Approximations. The
\texttt{gmvjoint}  \cite{murray2023package} is used for joint modeling of survival and multivariate longitudinal data. In this package, the longitudinal  model is specified using generalized linear mixed models. The joint models are fitted via maximum likelihood using the approximate EM algorithm developed by Bernhardt et al.  \cite{bernhardt2015fast}. The \texttt{jlctree} package  \cite{zhang2022joint} implements a tree-based approach for joint modeling of time-to-event and longitudinal data.
\texttt{jmBIG} offers Bayesian and non-Bayesian joint modeling for big data, particularly when dealing with very large sample sizes.
\texttt{VAJointSurv} is a software package that utilizes variational approximations to estimate joint modeling of multiple longitudinal markers and time-to-event outcomes. Each of these packages is provided for specific models and is based on certain assumptions.
For fitting different models, some other generic packages such as INLA  \cite{bivand2015spatial,bakka2018spatial}, rstan  \cite{guo2020package}, BayesX  \cite{brezger2005bayesx}, and BUGS  \cite{spiegelhalter2007openbugs} can be used to perform Bayesian inference.

The aim of this paper is to provide a comprehensive guide for readers interested in learning Bayesian inference for joint modeling of longitudinal measurements and time-to-event outcomes. The tutorial covers a range of models, starting from standard joint modeling, which includes a Gaussian longitudinal model and a  proportional hazard model, and progressing to more advanced techniques.
This paper also explores various methods of characterizing the relationship between longitudinal and time-to-event models. The joint modeling considers a proportional hazard sub-model with different forms of the baseline hazard and an accelerated failure time model. Additionally, it includes a discussion on multivariate longitudinal markers and competing risks. The discussion also includes more complex issues for the longitudinal data, such as zero-inflation. In all of the described models, the model is first introduced briefly. Then, the Bayesian paradigm is implemented, including model building and prior specification. Finally, the BUGS syntax for each model is provided. To implement the model, we start by simulating a dataset customized to the specifications of each model. Next, we utilize JAGS  \cite{plummer2012jags}, accessed seamlessly through R via the R2jags package, to fit the model to the simulated data. Finally, the model is deployed to analyze real datasets.
Although our focus is on running the code using the R2jags package, all codes can also be run independently in OpenBUGS or by calling them in R using R2OpenBUGS. The majority of the code shares a common structure for execution with both the R2OpenBUGS  \cite{sturtz2019r2openbugs} and R2jags  \cite{su2015package} packages. Any distinctions necessary to run the code with these packages are explicitly outlined. Additionally, Python, STATA, and Matlab interfaces for JAGS are now accessible, broadening the usefulness of this tutorial to users of these software programs as well.

This paper is structured as follows: Section 2 explores joint modeling with a single Gaussian longitudinal marker. Here, we outline the model specifications and simulate a corresponding dataset. We then apply the Joint Modeling (JM) approach, integrating a proportional hazard sub-model with various forms of baseline hazard  and various dependence structure between the event and the marker. We provide the BUGS syntax for each model.
In Section 3, we consider joint models involving multiple longitudinal markers. Section 4 centers joint models for competing risks, while Section 5 focuses on joint modeling of zero-inflated longitudinal markers.
Section 6 focuses on the joint modeling of longitudinal measurements along with time-to-event with cure fractions. Finally, we summarize our key findings and discuss advantages of the Bayesian approach
 in the concluding section.

\section{JM for one  longitudinal Gaussian marker   }
\subsection{General joint model formulation and specification}\label{s11}
Let $\mathcal{D}_n=\{T_i,\delta_i,{Y}_i;i=1,\ldots,n\}$ be a sample of $n$ subjects from the target population. let $T_i^*$ represent  the true time-to-event for subject $i$ and $C_i$ be the censoring time, $T_i$ is defined as the minimum of $T_i^*$ and $C_i$, and $\delta_i$ is the failure indicator defined as $I(T_i^*<C_i)$.
Let ${Y}_i$ be the $n_i$ vector of longitudinal responses for subject $i$, where $Y_i(s_{ij})=Y_{ij}$ represents the observed longitudinal measurements at time $s_{ij}$. In this section, we assume the following linear mixed-effects model for the longitudinal outcome:
\begin{eqnarray}\label{long}
Y_i(s_{ij})=\mu_i(s_{ij})+{\varepsilon}_{ij}={x}^\top_i(s_{ij}){\beta}+
{z}^\top_i(s_{ij}){b}_i+{\varepsilon}_{ij},
\end{eqnarray}
where ${x}_i(s)$ and ${z}_i(s)$ denote the time-dependent design vectors  with corresponding vectors  of fixed effects ${\beta}$ and  subject-specific random effects with ${b}_i\sim \mathcal{N}({0},{D})$. It is assumed that ${\varepsilon}_{ij}$s are independent and normally distributed with a mean vector of zero and a variance of $\sigma^2$ (${\varepsilon}_{ij}\sim \mathcal{N}(0,\sigma^2)$). Additionally, it is assumed that ${\varepsilon}_{ij}$ is independent of ${b}_i$. For the survival time-to-event outcome, we assume the following hazard function:
\begin{eqnarray}\label{surv}
\lambda_i\left(t \mid \mathcal{H}_i(t), {w}_i\right) =\lambda_0(t) \exp \left[{\alpha}^{\top} {w}_i+g\left(\mathcal{H}_i(t), {b}_i, {\gamma}\right )\right], \quad t>0,
\end{eqnarray}
where $\mathcal{H}_i(t)=\{\mu_i(s),~0<s\leq t\}$  is the history of the longitudinal process up to time t, $\lambda_0(.)$ is the baseline hazard function, while ${w}_i$ represents a vector of baseline covariates that correspond to the regression coefficients ${\alpha}$. Function $g(.)$ is a known function of $\mathcal{H}_i(t)$ and the random effects ${b}_{i}$ parameterized by a vector ${\gamma}$. This function specifies the features of the longitudinal outcome process that are included in the linear predictor of the relative risk model.\\
The most commonly used forms of $g(.)$ are 
 \cite{rizopoulos2011bayesian,rizopoulos2012joint,taylor2013real,rizopoulos2014combining,rizopoulos2016personalized}:
\begin{eqnarray}\label{l1}
g\left(\mathcal{H}_i(t), {b}_i, {\gamma}\right )&=&{\gamma} \mu_i(t), 
\end{eqnarray}
\begin{eqnarray}\label{2} g\left(\mathcal{H}_i(t), {b}_i, {\gamma}\right )&=&{\gamma} \mu_i^{\prime}(t), \quad \text { with } \mu_i^{\prime}(t)=\frac{\mathrm{d} \mu_i(t)}{\mathrm{d} t}, 
\end{eqnarray}
\begin{eqnarray}\label{3} g\left(\mathcal{H}_i(t), {b}_i, {\gamma}\right )&=&{\gamma} \int_0^t \mu_i(s) \mathrm{d} s, 
\end{eqnarray}
\begin{eqnarray}\label{4} g\left(\mathcal{H}_i(t), {b}_i, {\gamma}\right )&=&{\gamma}^{\top} {b}_i.
\end{eqnarray}
By considering formulation \eqref{l1}, we assume that the hazard of an event at time $t$ may be associated with the underlying level of the biomarker at the same time point. Based on equation \eqref{2}, the hazard is associated with the slope of the longitudinal profile at time $t$. By considering equations \eqref{3} or \eqref{4}, we assume that the hazard of an event at time $t$ may be associated with the accumulated longitudinal process up to time $t$ or the random effects, respectively  \cite{rizopoulos2016personalized}.
\subsection{Bayesian estimation}
For Bayesian implementation, we consider prior distributions for the unknown parameters, including regression coefficients and the variance of the errors. Specifically, we assume that ${\beta}$ follows a normal distribution, ${\beta}\sim \mathcal{N} ({\mu}_{\beta},{\Sigma}_{\beta})$ and $\sigma^2$ follows an inverse gamma distribution, $\sigma^2\sim \mathcal{IG} (a_\sigma,b_\sigma)$. Additionally, for the survival sub-model, we make the following assumption: ${\alpha}\sim \mathcal{N} ({\mu}_{\alpha},{\Sigma}_{\alpha})$ and ${\gamma}\sim \mathcal{N} ({\mu}_{\gamma},{\Sigma}_{\gamma})$. Also, ${D}\sim \mathcal{IW}({\Omega}_{D},\omega_{D})$. Let ${\theta}_{\lambda_0}$ be the vector of parameters for the baseline hazard, and let us to assume  ${\theta}_{\lambda_0} \sim \pi({\theta}_{\lambda_0})$. The choice of this prior depends on the structure of the baseline risk as discussed later. Then, given 
${y}^\top=({y}_1^\top,\ldots,{y}_n^\top)$, ${t}^\top=(t_1,\ldots,t_n)$, and ${\delta}^\top=(\delta_1,\ldots,\delta_n)$, the joint posterior distribution for the unknown parameters ${\theta}=({\beta},\sigma^2,{D},{\alpha},{\gamma},{\theta}_{\lambda_0})$ and the vector of random effects is given by:
\begin{eqnarray}\label{post}
\begin{aligned}
& 
\pi({\theta}, {b} \mid {y}, {t}, {\delta})  \propto \prod_{i=1}^n\left( \prod_{j=1}^{n_i} \phi\left(y_{i j };\mu_i(s_{ij}), \sigma^2\right) \right) \\
& \times \lambda_i\left(t_i \mid {w}_i, {b}_i\right)^{\delta_i} \exp \left(-\Lambda_i\left(t_i \mid {w}_i, {b}_i\right)\right) \\
& \times \phi\left({b}_i ; \mathbf{0}, {D}\right) \times \pi({\theta}),
& 
\end{aligned}
\end{eqnarray} 
where $\phi\left(.;\mu, \sigma^2\right)$ denotes a univariate normal distribution
with mean $\mu$ and variance $\sigma^2$, and $\Lambda_{i}(t_i|{\omega}_i,{b}_i)=\int_0^{t_i} \lambda_{i}(u|{\omega}_i,{b}_i)  du$. Under the assumption of independent priors, $\pi({\theta})$ is the product of priors for its components. Also, in this paper, the hyperparameters are chosen to result in non-informative priors.
Gaining precise Bayesian estimates through the utilization of the posterior distribution \eqref{post} is not practically achievable. Nevertheless, the BUGS syntax provides tools to conduct Bayesian analyses on complex statistical models using Markov chain Monte Carlo (MCMC) techniques.
In the following, we first describe how to simulate data from the joint models \eqref{long} and \eqref{surv}. Then, by adding some assumptions to fully specify the model, we implement Bayesian inference.
For more details on MCMC methods, particularly Gibbs sampling,  refer to Bolstad  \cite{bolstad2009understanding}.

\subsection{Data simulation}\label{simuni}
\begin{figure}
\centering
\includegraphics[width=10cm]{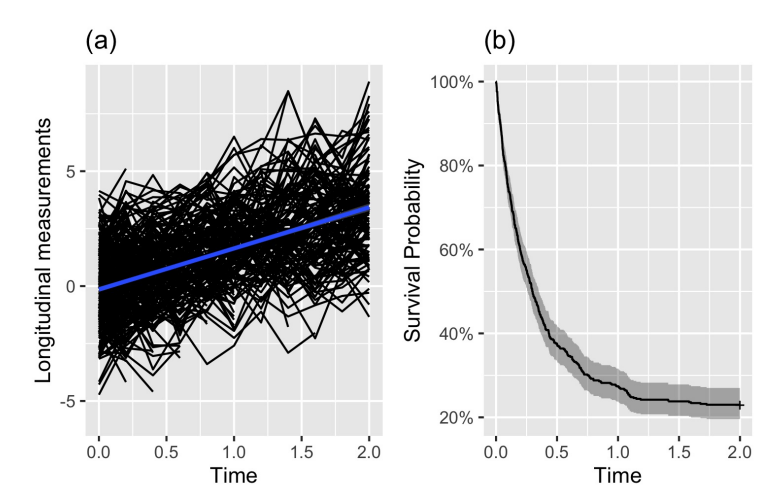}
\vspace*{-.1cm} \caption{\label{sim1}  
The individual trajectories for the generated longitudinal data and the Kaplan-Meier survival curve of time-to-event for the first set of the simulated data }
\end{figure}

In this section, we consider a JM with one Gaussian longitudinal marker. For this purpose, first, the process of generating the data for this model will be described. Then, the analysis of the simulated data will be discussed, utilizing JMs with different baseline hazards and different forms of association.
Data for $n=500$ subjects was simulated using a joint model with a Gaussian longitudinal marker and the following mixed effect model:
\begin{eqnarray}\label{gau}
Y_{i}(s_{ij})&=&\mu_{i}(s_{ij})+\varepsilon_{ij}\\
&=&\beta_{0}+\beta_{1}s_{ij}+\beta_2 x_{1i}+\beta_3 x_{2i}+b_{0i}+b_{1i} s_{ij}+\varepsilon_{ij},\nonumber
\end{eqnarray}
where $x_1$ is a binary covariate with a success probability of 0.6, and $x_2$ is a standard normal covariate and the marker measurements were generated at times $0,0.2,0.4,\ldots,2$.
$\varepsilon_{ij} \overset{\text{iid}}{\sim} \mathcal{N}(0, \sigma^2)$, ${b}_i=(b_{0i},b_{1i})^\top \overset{\text{iid}}{\sim} N_2({0},{D})$, where $i=1,\ldots,n$. Additionally, $\varepsilon_{ij}$ and ${b}_i$ are independent.
The time-to-event was simulated according to a proportional hazard model that depended on the current value of the longitudinal marker specified by:
\begin{eqnarray}\label{cox}
\lambda(t)=\lambda_0(t) \exp ({{\alpha}}^\top{w}_i+ \gamma \mu_{i}(t)).
\end{eqnarray}
The measurement of the longitudinal markers was generated by first drawing the individual random effects and the covariates from their distributions for each subject, and then adding the random errors for each time of measurement. Data generation from the time-to-event model with a time-dependent covariate \eqref{cox} is more complicated. The most common algorithm for simulating time-to-event is to use the inverse of the cumulative hazard function  \cite{austin2012generating} by employing the inverse probability integral transform  \cite{ross2012simulation}. This algorithm involves generating a random variable $U$ from a uniform distribution between 0 and 1 ($U(0,1)$) and solving the equation $U=S(t)$ for $t$. When the baseline hazard is constant and $\mu_i(t)$ is linear in \eqref{cox}, this equation has a closed-form solution. In other cases, a univariate root-finding algorithm, such as Brent's algorithm  \cite{brent2013algorithms} may be used. 
The package \texttt{simsurv} \cite{brilleman2021simulating} implements the algorithm developed by Crowther and Lambert  \cite{crowther2012simulating} that combines numerical integration to compute the cumulative hazard and numerical root finding.\\
 Here, we consider two different simulated datasets: the first one assumes a constant baseline hazard (i.e. $\lambda_0(t)=\lambda_0$), so the cumulative hazard function can be expressed in the following closed form.
\begin{eqnarray}\label{chaz}
\Lambda(t_i)=\frac{\lambda_0\exp({A_{0i}})}{A_{1i}}\big(\exp({A_{1i}t_i})-1\big),
\end{eqnarray}
where 
\begin{eqnarray}\label{aoa1}
~A_{0i}={{\alpha}}^\top{w}_i+\gamma(\beta_{0}+\beta_2 x_{1i}+\beta_3 x_{2i}+b_{0i}),~A_{1i}=\gamma(\beta_{1}+b_{1i}),
\end{eqnarray}
and the equation $u_i=S(t_i)$, where $u_i$ is a uniformly distributed random number between 0 and 1, can be easily solved by:
\begin{eqnarray}\label{ti}
t_i=\frac{1}{A_{1i}}\log(1-\frac{A_{1i}\log(u_i)}{\lambda_0 \exp(A_{0i})}).
\end{eqnarray}
It is important to note that the support of $A_{0i}$ is $\mathbb{R}$ and $A_{1i}$ is $\mathbb{R}-\{0\}$. The generation of the number for any values of $A_{0i}$ and $A_{1i}$ greater than 0 does not pose any problem. But for some values of $A_{1i}<0$ and especially when $A_{1i}\xrightarrow{}{-\infty}$, then $t_i\xrightarrow{} \infty$, and it should be considered as right censored. 
In general, the value of $t_i$ should not be negative. We have demonstrated that for certain parameter values and a given $u$ value, it is not possible to compute the solution. This situation arises when the $t_i$ values are very large, indicating censored time to event.
For the extension of this approach to more complex baselines, refer to Austin  \cite{austin2012generating}.
One can also use the \texttt{PermAlgo}  \cite{sylvestre2017permalgo}, \texttt{simjm} \cite{brilleman2018} and \texttt{simsurv} \cite{brilleman2021simulating} R packages as well as the \texttt{simulateJM} function of \texttt{JM} \cite{JM} or the {\it{simjoint}} function of \texttt{joineR} \cite{philipson2012joiner} for generating samples according to a  joint model.\\
{\bf{Example: Generation using equation \eqref{ti}}:} The real values of parameters are considered as 
$-\beta_0=\beta_1=\beta_2=\beta_3=0.5$, $\sigma^2=1$, ${s}_i=(0,0.2,0.4,\cdots,2)^\top$,
${D}=\begin{pmatrix}
1 & 0.5 \\
0.5 & 1 
\end{pmatrix}$, $\gamma=-0.5$ and ${{\alpha}}^\top{w}_i={{\alpha_1}}{w}_{1i}+{{\alpha_2}}{w}_{2i}$, where ${w}_{1i}$ is a standard normal covariate and 
${w}_{2i}$ is a binary covariate with a success probability of 0.5. 
The true event time $T_i^*$  was generated by \eqref{ti} and the censoring time $C_i$ was generated according to an exponential distribution with mean 0.5 plus an administrative censoring at time 2. The observed time $T_i=min(T_i^*, C_i)$
 represents the final survival time. Subsequently, the longitudinal repeated measurements after this time are considered as missing values.
The R codes and the simulated data can be found at
\url{https://github.com/tbaghfalaki/JM-with-BUGS-and-JAGS/tree/main/Simulated_data}.   Figure \ref{sim1} 
shows the individual trajectories for the generated longitudinal data and the Kaplan-Meier survival curve for survival time in the first set of the simulated data.\\

\subsection{JM estimation for different baseline hazard functions}
\subsubsection{Constant baseline hazard}
In this section, we demonstrate how to implement a joint model defined by \eqref{gau} and \eqref{cox} using BUGS, with a constant baseline hazard. For this purpose, we first consider the longitudinal sub-model, such that,
$Y_{ij}\sim \mathcal{N}(\mu_{ij},\sigma^2),$ where $\mu_{ij}=\beta_{0}+\beta_{1}t_{ij}+\beta_2 x_{1i}+\beta_3 x_{2i}+b_{0i}+b_{1i} t_{ij}={X}_{ij}^\top{\beta}+{Z}_{ij}^\top{b}_i,$ and ${X}_{ij}$ and ${Z}_{ij}$ represent the design matrices for the $i$th subject at measurement $j$. Also, $\tau^2=1/\sigma^2$ represents the precision parameter. For the longitudinal sub-model, we have the following BUGS code in Listing \ref{llll}.\\

\begin{lstlisting}[caption=The longitudinal sub-model in BUGS syntax.,label=llll]
for(i in 1:n){
    #Longitudinal sub-model
    for(j in 1:M[i]){
      Y[i,j]~dnorm(mu[i,j],tau)
      mu[i,j]<-inprod(beta[],X[i,j,])+inprod(b[i,],Z[i,j,])
    }
  }  
\end{lstlisting}
To handle the proportional hazard model in BUGS, it is necessary to use the zero-trick  \cite[p.276]{ntzoufras2011bayesian}, which is a well-known method for modeling non-standard distributions or those that are not available in BUGS. In this approach, a Poisson distribution can be used to indirectly specify the model likelihood. Let $L_i=f(t_i,\delta_i)$ and $\ell_i=\log(L_i)$ represent the likelihood and log-likelihood of the proportional hazard model, respectively. Then, this part of the likelihood   can be written as:
$$L_i \propto \exp(\ell_i-C)=\frac{\exp(-(C-\ell_i))(C-\ell_i)^0}{0!}=f_{Pos}(0;C-\ell_i),$$
where $f_{Pos}(0;C-\ell_i)$ is a Poisson distribution with a mean equal to $C-\ell_i$, and the observed value is set equal to zero. Also, $C$ is a positive constant term which must be selected in such a way that $C-\ell_i>0$, where the log-likelihood has the following form:
$$\ell_i={\delta_i}\log(\lambda(t_i))+ \log(S(t_i))={\delta_i}\log(\lambda(t_i))-\Lambda(t_i),$$
where $S(.)$ is the survival function and   $\Lambda(.)$ is given in equations \eqref{chaz} and \eqref{aoa1}.
A similar trick is also available as a one-trick, and for more details, refer to  Ntzoufras \cite{ntzoufras2011bayesian}.
The BUGS code for the joint model defined by \eqref{gau} and \eqref{cox} and with constant baseline hazard is given in Listing \ref{baselineconstant}. 
Note that in defining the baseline hazard, we make the assumption that $\log(\lambda_0)=\alpha_0$, where $\alpha_0$ represents the intercept of the survival model. We further specify $\alpha_0\sim N(0,1000)$ as the prior distribution for $\alpha_0$.
In this code, $C=100000$ is considered to ensure the 
positivity of the mean of the Poisson distribution. 
Also, one can find the full R codes of the JM with a proportional hazard sub-model with a constant baseline hazard in 
\url{https://github.com/tbaghfalaki/JM-with-BUGS-and-JAGS/tree/main/constant_BH}.
\begin{lstlisting}[caption=The JM with constanct baseline hazard in BUGS syntax.,label=baselineconstant]
model{
  for(i in 1:n){
    # Longitudinal sub-model
    for(j in 1:M[i]){
      Y[i,j]~dnorm(mu[i,j],tau)
      mu[i,j]<-inprod(beta[],X[i,j,])+inprod(b[i,],Z[i,j,])
    }
    A0[i]<- gamma*(beta[1]+beta[3]*X1[i]+beta[4]*X2[i]+b[i,1])
    A1[i]<- gamma*(beta[2]+b[i,2])
    # Survival sub-model
    # Hazard function
    haz[i]<- exp(inprod(alpha[],W[i,])+
                   (A0[i]+A1[i]*st[i]))
    # Cumulative hazard function
    chaz[i]<- exp(inprod(alpha[],W[i,])+A0[i])*(exp(st[i]*A1[i])-1)/A1[i]
    # Log-survival function
    logSurv[i]<- -chaz[i]
    # Definition of the survival log-likelihood using zeros trick
    phi[i]<-100000-death[i]*log(haz[i])-logSurv[i]
    zeros[i]~dpois(phi[i])
    # Random effects
    b[i,1:Nb]~dmnorm(mub[],Omega[,])
  }
  # Prior distributions
  for(l in 1:Nbeta){
    beta[l]~dnorm(0,0.001)
  }
  for(l in 1:Nalpha){
    alpha[l]~dnorm(0,0.001)
  }
  gamma~dnorm(0,0.001)
  tau~dgamma(0.01,0.01)
  Omega[1:Nb,1:Nb]~dwish(V[,],Nb)
  # Determinetic nodes
  lambda0<-exp(alpha[1])
  sigma<-1/tau
  Sigma[1:Nb,1:Nb]<-inverse(Omega[,])
}
\end{lstlisting}
\FloatBarrier
\subsubsection{Weibull baseline hazard }\label{this}
For a parametric proportional hazard model with a Weibull baseline hazard, we consider the following hazard function:
\begin{eqnarray}\label{wei}
\lambda(t)=\lambda_0 \nu t^{\nu-1} \exp ({{\alpha}}^\top{w}_i+ \gamma\mu_{i}(t)).
\end{eqnarray}
As in the preceding sub-section, the zero-trick method is employed as a solution for model building. However, since the cumulative hazard function for this model does not have a closed form, $\Lambda(t)$ is calculated using either the Gauss–Legendre or Gauss–Kronrod quadrature methods (refer to Supplementary Material A for more details). For this purpose, first we define the nodes and corresponding weights for Gauss-Legendre quadrature or Gauss-Kronrod quadrature (the R codes for these Gaussian quadratures are given in Listings A. \\
The cumulative hazard function of equation \eqref{wei} can be approximated by:
$$\Lambda(t)\approx \lambda_0 \nu \sum_{k=1}^K \tilde w_k(t) \tilde x_k(t)^{\nu-1} \exp ({{\alpha}}^\top{w}_i+\gamma \mu_{i}(\tilde x_k(t))),$$
where $\tilde x_k(t)=\frac{x_k+1}{2} t$ and $\tilde w_k(t)=\frac{w_k}{2} t$ are the scaled values of the node $x_k$ and the weight $w_k$, respectively. 
Furthermore, the prior distribution for $\nu$ is specified as $\Gamma(0.01,0.01)$.
Thus, this model is estimated by replacing lines 10-17 in listing \ref{baselineconstant} with the code in listing \ref{bugswei}. The complete R code for the JM with a proportional hazard sub-model and a Weibull baseline hazard is provided in \url{https://github.com/tbaghfalaki/JM-with-BUGS-and-JAGS/tree/main/Weibull_BH}.
\begin{lstlisting}[caption=The survival sub-model for the weibull model in BUGS syntax.,label=bugswei]
    # Hazard function 
    haz[i]<- nu*pow(st[i],nu-1)*exp(inprod(alpha[],W[i,])+
                                            (A0[i]+A1[i]*st[i]))
    #Cumulative hazard function 
    
    for(j in 1:K){
      # Scaling Gauss-Kronrod/Legendre quadrature 
      xk1[i,j]<-(xk[j]+1)/2*st[i] 
      wk1[i,j]<- wk[j]*st[i]/2
      #  Hazard function at Gauss-Kronrod/Legendre nodes
      chaz[i,j]<- nu*pow(xk1[i,j],nu-1)*exp(inprod(alpha[],W[i,])+
                                                    (A0[i]+A1[i]*xk1[i,j]))
    }
    
    #Log-survival function with Gauss-Kronrod/Legendre requadrature
    logSurv[i]<- -inprod(wk1[i,],chaz[i,])
\end{lstlisting}
\subsubsection{Piecewise constant baseline hazard}
One of the most popular baseline hazard functions is the piecewise constant baseline \cite{ibrahim2001bayesian}. In this model, a finite partition of the time axis can be constructed initially as $0=s_0<s_1<s_2<\ldots,<s_J$ such that $max(t_1,\ldots,t_n)<s_J$. Thus, there are $J$ intervals such as $(0,s_1],(s_1,s_2],\ldots,(s_{J-1},s_J]$. We consider a constant baseline hazard for each interval as $\lambda_0(t)=h_j,~t\in (s_{j-1},s_j],~j=1,\ldots,J$.
The cumulative hazard function for $t_i<s_1$ is given by:
\begin{eqnarray}
\Lambda_i(t) &=& \frac{\exp(A_{0i})}{A_{1i}}h_1(\exp(A_{1i}t)-1),
\end{eqnarray}
for $s_{j-1}<t\leq s_j,~j>1$, we have
\begin{eqnarray}
\Lambda_i(t) &=& \frac{\exp(A_{0i})}{A_{1i}}\Big(h_j(\exp(A_{1i}t)-\exp(A_{1i}s_{j-1}))+\sum_{k=1}^{j-1}h_k(\exp(A_{1i}s_k)-\exp(A_{1i}s_{k-1})) \Big),
\end{eqnarray}
where $A_{0i}$ and $A_{1i}$ are given in equation \eqref{aoa1}. 
A common choice for the prior distribution for $h_j,~j=1,\ldots,J$ is independent gamma priors which are denoted as $h_j\sim \mathcal{G} (\nu_{0j},\nu_{1j})$. The values of the hyper-parameters $\nu_{0j}$ and $\nu_{1j}$ can be elicited in a way that leads to a low-informative prior. Other priors for $h_j,~j=1,\ldots,J$ are considered by Ibrahim et al.  \cite{ibrahim2001bayesian,sinha1993semiparametric,leonard1978density} and Christensen et al.  \cite{christensen2010bayesian}.\\
Listing \ref{piecewise} shows the BUGS code for the hazard and cumulative hazard functions for a piecewise baseline hazard with $J=5$ pieces. The complete R code for the JM with a proportional hazard sub-model with a piecewise constant baseline hazard can be found at
\url{https://github.com/tbaghfalaki/JM-with-BUGS-and-JAGS/tree/main/PC_BH}.

\begin{lstlisting}[caption=The hazard and cumulative hazard for piecewise baseline hazard in BUGS syntax.,label=piecewise]
  A0[i]<- inprod(alpha[],W[i,])+gamma*(beta[1]+beta[3]*X1[i]+beta[4]*X2[i]+b[i,1])
    A1[i]<- gamma*(beta[2]+b[i,2])
    #Survival sub-model
    #Hazard function
    haz[i]<- ((h[1]*step(s[1]-st[i]))+
             (h[2]*step(st[i]-s[1])*step(s[2]-st[i]))+
             (h[3]*step(st[i]-s[2])*step(s[3]-st[i]))+
             (h[4]*step(st[i]-s[3])*step(s[4]-st[i]))+
             (h[5]*step(st[i]-s[4])))*exp(A0[i]+A1[i]*st[i])
  
      #Cumulative hazard function
    chaz1[i]<- h[1]*(exp(A1[i]*st[i])-1)*step(s[1]-st[i])+
      (h[1]*(exp(A1[i]*s[1])-1)+h[2]*(exp(A1[i]*st[i])-exp(A1[i]*s[1])))*(step(st[i]-s[1])*step(s[2]-st[i]))+
      (h[1]*(exp(A1[i]*s[1])-1)+h[2]*(exp(A1[i]*s[2])-exp(A1[i]*s[1]))+h[3]*(exp(A1[i]*st[i])-exp(A1[i]*s[2])))*(step(st[i]-s[2])*step(s[3]-st[i]))+
      (h[1]*(exp(A1[i]*s[1])-1)+h[2]*(exp(A1[i]*s[2])-exp(A1[i]*s[1]))+h[3]*(exp(A1[i]*s[3])-exp(A1[i]*s[2]))+h[4]*(exp(A1[i]*st[i])-exp(A1[i]*s[3])))*(step(st[i]-s[3])*step(s[4]-st[i]))+
      (h[1]*(exp(A1[i]*s[1])-1)+h[2]*(exp(A1[i]*s[2])-exp(A1[i]*s[1]))+h[3]*(exp(A1[i]*s[3])-exp(A1[i]*s[2]))+
      h[4]*(exp(A1[i]*s[4])-exp(A1[i]*s[3]))+h[5]*(exp(A1[i]*st[i])-exp(A1[i]*s[4])))*step(st[i]-s[4])

    chaz[i]<-exp(A0[i])*chaz1[i]/A1[i]
    #Log-survival function
    logSurv[i]<- -chaz[i]
\end{lstlisting}

\subsection{Spline baseline hazard }
Another approach for modeling the baseline hazard function is to use a B-spline approach. Let a set of equally spaced knots within the domain of ${t}=(t_1,\ldots,t_n)^\top$ be selected. In particular, the logarithm of the baseline hazard function is expressed as follows.
\begin{eqnarray}\label{d1}
\log(\lambda_0(t))
=\alpha_{\lambda_0,0}+\sum_{l=1}^L \alpha_{\lambda_0,l} \mathcal{B}_l(t,{\nu},\kappa_1),
\end{eqnarray}
where $\mathcal{B}_l(t,{\nu},\kappa_1)$ represents the $l$th basis function of a B-spline with a degree of $\kappa_1$ and knots $\nu_1,\ldots,\nu_{\kappa_2}$. Here, $L$ is defined as $\kappa_1+\kappa_2$. Additionally, ${\alpha}_{\lambda_0}=(\alpha_{\lambda_0,0},\ldots,\alpha_{\lambda_0,L})^\top$ denotes the vector of spline coefficients.\\
In the frequentist approach, this model can be estimated by maximizing the penalized likelihood with the following penalty:
\begin{eqnarray}\label{d2}
- \tau \sum_{l=m+1}^L\big(\Delta^{(m)}\alpha_{\lambda_0,l}\big)^2 ,
\end{eqnarray}
where the notation $\Delta^{(m)}$ represents the $m$th order backward difference operator. The value of $m$ is often chosen to be 1 or 2, which leads to linear (degree = 1) and cubic (degree = 3) smoothing splines, respectively.
Under the Bayesian paradigm, we replace the difference penalties in \eqref{d2} by their stochastic counterparts  \cite{lang2004bayesian,speckman2003fully}. The first difference corresponds to a first-order random walk, and the second difference corresponds to a second-order random walk as follows:
\begin{eqnarray}\label{d3}
u_{j,\rho}=\alpha_{\lambda_0,j,\rho}-\alpha_{\lambda_0,j,\rho-1},~~
u_{j,\rho}=\alpha_{\lambda_0,j,\rho}-2\alpha_{\lambda_0,j,\rho-1}+\alpha_{\lambda_0,j,\rho-2},
\end{eqnarray}
such that $u_{j,\rho}\sim \mathcal{N}(0,\sigma_j^2)$,  and thus ${\alpha}_{\lambda_0}$ follows a singular multivariate normal distribution with the following density function
\begin{eqnarray}\label{d4}
p({\alpha}_{\lambda_0}|\tau_{\lambda_0})\propto \exp\big(-\frac{\tau_{\lambda_0}}{2} 
{\alpha}_{\lambda_0}^\top {\Psi}_m{\alpha}_{\lambda_0}\big),
\end{eqnarray}
here $\tau_{\lambda_0}$ is the smoothing parameter that takes a $\mathcal{G}(1,0.005)$  hyper-prior in order to ensure a proper posterior for ${\alpha}_{\lambda_0}$ \cite{hobert1996effect,lang2004bayesian}, 
and ${\Psi}_m={K}_m^\top{K}_m$  for $m=1$ and $2$ are given as follows \cite{lang2004bayesian,wang2018bayesian,JSSv072i07}:
$$ {K}_1=\begin{pmatrix}
-1 & 1 &  &  &  \\
 & -1 & 1 &  &  \\
 &  & \ddots & \ddots &  \\
 &  &  & -1 & 1 
\end{pmatrix}, ~~{K}_2= \begin{pmatrix}
1 & -2 & 1 &  &  &  \\
 & 1 & -2 & 1 &  &  \\
 &  & \ddots & \ddots & \ddots &  \\
 &  &  & 1 & -2 & 1 
\end{pmatrix}.$$
For Bayesian inference, similar techniques as those discussed in Section \ref{this} can be employed, such as the zero trick and Gaussian quadrature, which can be beneficial. In this scenario, the cumulative hazard function for the baseline hazard \eqref{d1}, using $\tilde x_k(t)$ and $\tilde w_k(t)$, can be approximated as follows:
$$\Lambda(t)\approx \exp(
\alpha_{\lambda_0,0}+{{\alpha}}^\top{w}_i)
 \sum_{k=1}^K \tilde w_k(t)  \exp (
 \sum_{l=1}^L \alpha_{\lambda_0,l} \mathcal{B}_l(\tilde x_k(t),{\nu})+
\gamma \mu_{i}(\tilde x_k(t))).$$
To compute this quantity, we first calculate the base function of the B-spline for the values of the scaled Gauss-Legendre quadrature. Then, we import them into JAGS to perform Bayesian parameter estimation (see Listing \ref{spline}). Listing \ref{spline1} contains the BUGS code for the hazard and cumulative hazard functions of a spline baseline hazard. Also, the complete R codes for the JM with this baseline hazard can be found at \url{https://github.com/tbaghfalaki/JM-with-BUGS-and-JAGS/tree/main/Codes_3_5}.

\begin{lstlisting}[caption=R code for the base function of the B-spline for the values of the scaled Gauss-Legendre quadrature.,label=spline]
XK_sc <- matrix(0, K, n)
for (i in 1:n) {
  XK_sc[, i] <- (xk + 1) / 2 * st[i]
}
XK_t <- array(0, c(K, dim(st_bs)[2], n))
for (i in 1:n) {
  knots <- quantile(XK_sc[, i], prob = seq(0.1, 0.9, 0.15)) # B-spline
  XK_t[, , i] <- bSpline(XK_sc[, i], knots = knots, degree = 4, intercept = TRUE)
}
\end{lstlisting}

\begin{lstlisting}[caption=The hazard and cumulative hazard for the
proportional hazard model with a spline baseline hazard in BUGS syntax.,label=spline1]
    A0[i]<- inprod(alpha[],W[i,])+gamma*(beta[1]+beta[3]*X1[i]+beta[4]*X2[i]+b[i,1])
    A1[i]<- gamma*(beta[2]+b[i,2])
    # Survival sub-model
    # Hazard function 
    haz[i]<- exp(inprod(st_bs[i,],gamma_bs[])+A0[i]+A1[i]*st[i])
    # Cumulative hazard function 
    for(j in 1:K){
      # Scaling Gauss Kronrod (see wikipedia)
      x_scaled[i,j]<-(xk[j]+1)/2*st[i] 
      w_scaled[i,j]<- wk[j]*st[i]/2
      chaz[i,j]<-exp(inprod(XK_t[j,,i],gamma_bs[])+A0[i]+A1[i]*x_scaled[i,j])
    }
    # Log-survival function 
    logSurv[i]<- -inprod(w_scaled[i,],chaz[i,])
\end{lstlisting}

\subsection{Alternative dependence structures}
\subsubsection{Current value and current slope}
The code for considering associations based on the current slope is the same as that used for the current value. We are considering a proportional hazard model with the current value and current slope as follows:
\begin{eqnarray}\label{cox2}
\lambda(t)= \lambda_0(t)\exp ({{\alpha}}^\top{w}_i+ \gamma_1 \mu_{i}(t)+ \gamma_2 \mu_{i}^\prime(t)).
\end{eqnarray}
As an example, we consider the Weibull baseline hazard, that is, $\lambda_0(t)= \nu t^{\nu-1}$.
Of course, this association can be considered with different baseline hazard functions.
With $\mu_{i}(t)$ defined by equation \eqref{gau}, it is clear that
$$\mu_{i}^\prime(t)=\beta_1+b_{1i}.$$
The non-informative prior for $\gamma_2$ can be considered the same as that for  $\gamma_1$, which is $\gamma_2\sim \mathcal{N}(0,1000)$. Thus, in this status $A_{0i}$ in equation \eqref{aoa1} is replaced by 
\begin{eqnarray*}
A_{0i}={{\alpha}}^\top{w}_i+\gamma_1(\beta_{0}+\beta_2 x_{1i}+\beta_3 x_{2i}+b_{0i})+\gamma_2(\beta_{1}+b_{1i}).
\end{eqnarray*}
Thus, in Listing \ref{baselineconstant}, line 8 must be replaced by the following code:
\begin{lstlisting}
A0[i]<- gamma[1]*(beta[1]+beta[3]*X1[i]+beta[4]*X2[i]+b[i,1])+gamma[2]*(beta[2]+b[i,2])
\end{lstlisting}
The complete R code for considering both the current value and current slope can be found at \url{https://github.com/tbaghfalaki/JM-with-BUGS-and-JAGS/tree/main/current_slope}.
\subsubsection{Shared random effects }
The use of shared random effects is a simple approach to consider the association between longitudinal markers and survival outcomes  \cite{guo2004separate}. This approach allows us to use a standard survival model without time-varying covariates. The longitudinal model for the shared random effects remains unchanged [see equation \eqref{long}]. The time-to-event model will be simplified as follows:
\begin{eqnarray}\label{cox1}
\lambda(t)=\lambda_0(t) \exp ({{\alpha}}^\top{w}_i+ {{\gamma}}^\top{b}_i).
\end{eqnarray}
Thus, the cumulative hazard is given by
\begin{eqnarray}\label{cox1c}
\Lambda(t)= \Lambda_0(t)\exp ({{\alpha}}^\top{w}_i+ {{\gamma}}^\top{b}_i),
\end{eqnarray}
where $\Lambda_0(t)=\int_0^t \lambda_0(t) du$. In the following example, we consider a Weibull baseline hazard with the function $\Lambda_0(t)=\lambda_0 t^\nu \exp ({{\alpha}}^\top{w}_i+ {{\gamma}}^\top{b}_i)$. The hazard, cumulative hazard, and logarithm of the survival function of Listing \ref{bugswei} (lines 1-13) should be replaced with Listing \ref{shared1} to obtain the results for this joint model.
\begin{lstlisting}[caption=The hazard and cumulative hazard for shared random effects model in BUGS syntax.,label=shared1]
 haz[i]<- nu*pow(st[i],nu-1)*exp(inprod(alpha[],W[i,])+inprod(b[i,],gamma[]))
    #Cumulative hazard function 
    chaz[i]<- pow(st[i],nu)*exp(inprod(alpha[],W[i,])+inprod(b[i,],gamma[]))
    #Log-survival function with Gauss-Kronrod/Legendre requadrature
    logSurv[i]<- -chaz[i]
## prior for association parameters
for(l in 1:2){ 
  gamma[l]~dnorm(0,0.001)
}
\end{lstlisting}
While the Weibull model in both Accelerated Failure Time (AFT) models and the proportional hazard model produces identical outcomes, it is noteworthy that when considering the shared random effect model in BUGS, using AFT models is simpler than opting for the proportional hazard model.
 For this purpose, we will consider the Weibull model. However, the details remain the same for other members of the AFT models.
The Weibull distribution with parameters $(v_1,v_2)$ [we use the notation $T\sim Weibull (v_1,v_2)$ to denote this distribution] in BUGS syntax is defined as $v_1 v_2 t^{v_1-1}\exp(-v_2t^{v_1}),~t>0$. For the Weibull model, we assume that $T_i$ follows a Weibull distribution with parameters $v_1$ and $\exp ({{\alpha}}^\top{w}_i+ {{\gamma}}^\top{b}_i)$. \\
JAGS employs the $\texttt{dinterval}(,)$ function to model censored data within a distribution. This function is parameterized by two values, 0 and 1. When expressed as $\texttt{is.censored}[i]\sim \texttt{dinterval}(t[i], c[i])$, it signifies that for censored data, $\texttt{is.censored}[i]=1$, whereas for observed data, $\texttt{is.censored}[i]=0$.\\
Survival analysis requires the definition of pairs $t[i]$ and $c[i]$. In this context, $t[i]$ denotes either an actual time or a missing value (=NA), depending on whether the observation corresponds to the time of the actual event or a censoring time, respectively. Additionally, $c[i]$ represents the censoring time. In instances where the $i$th individual is not censored, a substantially large number is assigned to $c[i]$. This large value often symbolizes the longest possible follow-up time in the study being analyzed \cite{rosner2021bayesian}. For clarity on implementing these distinctions between observed and censored data in JAGS, refer to the R syntax provided in Listing \ref{censjags}.
\begin{lstlisting}[caption=The right censored data in R syntax for JAGS.,label=censjags]
is.censored=1-death
t=c=rep(0,n)
for(i in 1:n){
  if(death[i]==0)((t[i]=NA) & (c[i]=st[i]))
  if(death[i]==1)((t[i]=st[i]) & (c[i]=max(st)))
}
\end{lstlisting}
For more details about interval censoring or left censoring, refer to Rosner et al. \cite{rosner2021bayesian} and Alvares et al. \cite{alvares2021bayesian}. After preparing the data, the BUGS code for this purpose is provided in Listing \ref{jagscode}.
\begin{lstlisting}[caption=The BUGS code for the AFT model for JAGS.,label=jagscode]
t[i] ~ dweib(kappa,mut[i]) 
    log(mut[i])<-inprod(alpha[],W[i,])+inprod(b[i,],gamma[])
    is.censored[i]~dinterval(t[i],c[i])
\end{lstlisting}
It should be mentioned that specifying censoring in OpenBUGS and JAGS differs. For more details about 
this difference refer to \url{https://github.com/tbaghfalaki/JM-with-BUGS-and-JAGS/blob/main/shared_RE/README.md}. 
All the R codes including a set of generated data under the assumption of the shared random effects model can be accessed at \url{https://github.com/tbaghfalaki/JM-with-BUGS-and-JAGS/tree/main/shared_RE}. \\
This section involves simulating data and analysing them from a single Gaussian longitudinal marker and time-to-event outcome. To see an illustrative of real application  of such a dataset, one may  refer to  \cite{alvares2021bayesian}.
In the upcoming sections, we will expand on joint modeling. We will use both simulated data and real datasets to illustrate each extension.

 \section{JM for multiple longitudinal markers }
\subsection{Model specification}
Consider $\mathcal{D}_n=\{T_i,\delta_i,{y}_{ik};i=1,\ldots,n,k=1,\ldots,K\}$ as a sample of $n$ individuals from the target population. Also, let ${y}_{ik}$ be the $n_{ik}$ vector of longitudinal responses for subject $i$ and the $k$th marker, where $y_{ijk}$ represents the observed longitudinal measurements at time $s_{ijk}$. We assume a generalized linear mixed-effects model for the longitudinal outcome, where the distribution of $Y_{ijk}$ is represented by the following form:
\begin{equation}
f_{Y_{ijk}}(y)=\exp{ \{ \psi_k^{-1} [\theta_{ijk} y - h_k(\theta_{ijk})] +C_k(y,\psi_k) \}},
\label{density}
\end{equation}
where  $\theta_{ijk}$ is often referred to as the canonical parameter, and $h_k(.)$ and $C_k(.,.)$ are known functions. The formulation of the generalized linear mixed model is completed by the following assumption:
\begin{eqnarray}\label{mixed}
E(Y_{ik}(s)|{b}_{ik})= h_k'(\theta_{ijk})
 = h_k'({x}^\top_{ik}(s){\beta}_k+
{z}^\top_{ik}(s){b}_{ik}),
\end{eqnarray}
where the link function $h_k'(.)$ in (\ref{mixed}) represents the derivative of the function $h_k(.)$ in (\ref{density}); ${x}_{ik}$ and ${z}_{ik}$ represent vectors of covariates. These covariates may include time-dependent exogenous covariates, the time variable $s_{ijk}$ and interactions with $s_{ijk}$, and ${z}_{ik} \subset {x}_{ik}$.
The vector of random effects is ${b}_i=({b}_{i1},\ldots,{b}_{iK})$, and ${b}_i\sim \mathcal{N}({0},{D})$. For the time-to-event model, we assume the following hazard function:
\begin{eqnarray}\label{hazzzzzz}
\lambda_i\left(t \mid \mathcal{H}_i(t), {w}_i\right) =\lambda_0(t) \exp \left[{\alpha}^{\top} {w}_i+\sum_{k=1}^K g_k\left(\mathcal{H}_{ik}(t), {b}_{ik}, {\gamma}_k\right )\right], \quad t>0,
\end{eqnarray}
where $\mathcal{H}_{ik}(t)=\{\mu_{ik}(s),~0<s\leq t\}$ is the history of the $k$th longitudinal process up to time $t$, and $\mu_{ik}(s)$ is the linear predictor in \eqref{mixed}, specifically, $\mu_{ik}(s)={x}^\top_{ik}(s){\beta}_k+
{z}^\top_{ik}(s){b}_{ik}$. 
The prior specification and Bayesian implementation of this JM are the same as those discussed in Section \ref{s11}.
\subsection{Codes for model estimation}
We present the code for estimating a joint model with $K=5$ longitudinal markers, including four Gaussian markers and one binary marker.
As an illustrative example, a data set was generated using the following model for the binary marker:
\begin{eqnarray}\label{bin}
{\rm{logit}} \left( P(Y_{i1}(s)=1| b_{01i},b_{11i} ) \right)&=&
\mu_{i1}(s)\\ &=&\beta_{01}+\beta_{11}s+\beta_{2k} x_{1i}+\beta_{3k} x_{2i}+b_{i10}+b_{i11}s, \nonumber
\end{eqnarray}
and for the four Gaussian markers the following model is used:
\begin{eqnarray}\label{gau2}
Y_{ik}(s)&=&\mu_{ik}(s)+\varepsilon_{ikt}\\
&=&\beta_{0k}+\beta_{1k}s+\beta_{2k} x_{1i}+\beta_{3k} x_{2i}+b_{ik0}+b_{ik1} s+\varepsilon_{ijk},~~k=2,\ldots,5, \nonumber
\end{eqnarray}
with $\varepsilon_{ijk} \sim \mathcal{N}(0, \sigma_k^2)$ and
where ${b}_i=(b_{i10},b_{i11},\ldots,b_{iK0},b_{iK1})^{\top} \sim \mathcal{N}({0},{D})$, $i=1,\cdots,n$ and $s\in \{0,0.2,0.4,\cdots,2\}$.\\
We assume that $g_k\left(\mathcal{H}_{ik}(t), {b}_{ik}, {\gamma}_k\right )=\mu_{ik}(t)$.
Thus, the time-to-event sub-model is specified by
$$\lambda(t)=\lambda_0(t) \exp (\alpha_1 x_1+\alpha_2 x_2+\sum_{k=1}^K \gamma_k \mu_{ik}(t)).$$
The dataset was generated utilizing a permutational algorithm proposed by Abrahamowicz et al.  \cite{abrahamowicz1996time} and Sylvestre and Abrahamowicz  \cite{sylvestre2008comparison}. This algorithm is specifically designed for generating data from the proportional hazard model, accounting for time-dependent covariates.  
The real values and the R code for generating data can be found at the following URL: \url{https://github.com/tbaghfalaki/JM-with-BUGS-and-JAGS/tree/main/multivariate}.\\
In Section 2, we consider various forms of baseline hazard for data analysis. Here, for simplicity, we consider a proportional hazard sub-model with a constant and a Weibull baseline hazard. However, one can consider alternative forms, similar to those discussed in Section 2. The BUGS code for the multi-marker JM is provided in Listing \ref{mm}, which includes a mixed effects logistic model for $Y_1$ and mixed effects models for the other markers.
\begin{lstlisting}[caption=The BUGS code for the multi-marker longitudinal outcomes.,label=mm]
for(j in 1:M[i]){
      Y1[i,j]~dbern(p1[i,j])
      logit(p1[i,j])<- mu1[i,j]
      mu1[i,j]<-inprod(betaL1[],XL1[i,j,])+inprod(b[i,1:2],ZL1[i,j,])
      Y2[i,j]~dnorm(mu2[i,j],tau[1])
      mu2[i,j]<-inprod(betaL2[],XL2[i,j,])+inprod(b[i,3:4],ZL2[i,j,])
      Y3[i,j]~dnorm(mu3[i,j],tau[2])
      mu3[i,j]<-inprod(betaL3[],XL3[i,j,])+inprod(b[i,5:6],ZL3[i,j,])
      Y4[i,j]~dnorm(mu4[i,j],tau[3])
      mu4[i,j]<-inprod(betaL4[],XL4[i,j,])+inprod(b[i,7:8],ZL4[i,j,])
      Y5[i,j]~dnorm(mu5[i,j],tau[4])
      mu5[i,j]<-inprod(betaL5[],XL5[i,j,])+inprod(b[i,9:10],ZL5[i,j,])
    }
\end{lstlisting}
The BUGS code for the survival sub-model is the same as that in Section 2 and is based on the zero-trick. However, the values of A0[i] and A1[i] are provided in Listing \ref{mm1}. Also, the complete R code is available at \url{https://github.com/tbaghfalaki/JM-with-BUGS-and-JAGS/tree/main/multivariate}.
\begin{lstlisting}[caption=The BUGS code for the survival sub-model.,label=mm1]
A0[i]<- inprod(betaS[],W[i,])+gamma[1]*(betaL1[1]+b[i,1])+gamma[2]*(betaL2[1]+b[i,3])+
      gamma[3]*(betaL3[1]+b[i,5])+gamma[4]*(betaL4[1]+b[i,7])+gamma[5]*(betaL5[1]+b[i,9])
A1[i]<- gamma[1]*(betaL1[2]+b[i,2])+gamma[2]*(betaL2[2]+b[i,4])+gamma[3]*(betaL3[2]+b[i,6])+
    gamma[4]*(betaL4[2]+b[i,8])+gamma[5]*(betaL5[2]+b[i,10])
\end{lstlisting}

\subsection{Application to PBC2 data}\label{pbc2_sec}
In this section, we analyze a real data set called PBC2  \cite{lin2002modeling} and provide guidance on using R software to confirm the convergence of results obtained from JAGS.\\
The dataset includes 312 patients who were enrolled in clinical trials at the Mayo Clinic from 1974 to 1984. The average follow-up time was 8.19 years, with a median of 5 visits and a maximum of 16 visits. In this analysis, the event of interest is death without transplantation, while subjects who are alive at the end of the study or have been transplanted are considered right-censored ($44.87\%$ dead). The complete data can be found in many R-packages, such as \texttt{joineRML}, which provides a description at the following URL: \url{https://rdrr.io/cran/joineRML/man/pbc2.html}. We consider five Gaussian biological markers, including albumin (in mg/dL), logarithm transformation of alkaline (alkaline phosphatase (in U/L), logarithm transformation of SGOT (in U/mL), logarithm transformation of platelets (platelets per cubic mL/1000), and logarithm transformation of serum bilirubin (serum bilirubin in mg/dl).\\
The sub-model for the time to death depends on the current values of the included markers and is adjusted for the treatment group (drug) and the patient's standardized age at enrollment. The change over time of the markers is described using linear mixed models, which include a linear time trend with random intercepts and random slopes. \\
We consider the following longitudinal sub-model:
\begin{eqnarray}\label{eq}
Y_{ik}(t)=\mu_{ik}(t)={X}_{ik}^T(t){\beta}_k+{Z}_{ik}^T(t){b}_{ki}+\varepsilon_{ik},~i=1,\cdots,n,~k=1,\cdots,5,
\end{eqnarray}
where $Y_1 = \rm{Albumin},~Y_2 = \log(\rm{Alkaline}), Y_3 = \log(\rm{SGOT}), ~Y_4 = \log(\rm{Platelets})$ and $Y_5 = \log(\rm{Serum~bilirubin})$. Also, ${X}_{ik}$ is a $n_i\times 2$ matrix such that the first column is equal to 1 for the random intercept and the second column is observed time. We consider random intercept and random slope for equation \eqref{eq}, therefore, ${Z}_{ik}={X}_{ik}$. Also,
${\varepsilon}_{ik}=(\varepsilon_{i1k},\cdots,\varepsilon_{in_{i}k})^\top$, such that, $\varepsilon_{ijk} \stackrel{ind}{\sim} \mathcal{N}(0, \sigma_k^2)$. For the survival sub-model, the hazard function is:
\begin{eqnarray}\label{coxnew}
\lambda(t)=\lambda_0(t) \exp (\alpha_0+\alpha_1 {\rm{Treat}}_i+\alpha_2 {\rm{Age}}_i+ \sum_{k=1}^5 \gamma_k \mu_{ik}(t)),
\end{eqnarray}
where $\lambda_0(t)=\nu t^{\nu-1}$. Also, ${b}_i=(b_{01i},b_{11i},b_{02i},b_{12i},\cdots,b_{05i},b_{15i})^{\top} \sim \mathcal{N}({0},{\Sigma})$.
The R code for preparing the data can be found in \url{https://github.com/tbaghfalaki/JM-with-BUGS-and-JAGS/blob/main/PBC/data_preparation_and_jags.R}. The BUGS code for analysing these data can be found in Listing \ref{pbc1}.

\begin{lstlisting}[caption=The BUGS code for analysing PBC2 data.,label=pbc1]
model{
  for(i in 1:n){
    #Longitudinalobservations
    for(j in 1:M[i]){
      Y1[i,j]~dnorm(mu1[i,j],tau[1])
      mu1[i,j]<-inprod(betaL1[],XL1[i,j,])+inprod(b[i,1:2],ZL1[i,j,])
      Y2[i,j]~dnorm(mu2[i,j],tau[2])
      mu2[i,j]<-inprod(betaL2[],XL1[i,j,])+inprod(b[i,3:4],ZL1[i,j,])
      Y3[i,j]~dnorm(mu3[i,j],tau[3])
      mu3[i,j]<-inprod(betaL3[],XL1[i,j,])+inprod(b[i,5:6],ZL1[i,j,])
      Y4[i,j]~dnorm(mu4[i,j],tau[4])
      mu4[i,j]<-inprod(betaL4[],XL1[i,j,])+inprod(b[i,7:8],ZL1[i,j,])
      Y5[i,j]~dnorm(mu5[i,j],tau[5])
      mu5[i,j]<-inprod(betaL5[],XL1[i,j,])+inprod(b[i,9:10],ZL1[i,j,])
    }
    #Survival and censoring times
    #Hazard function
    A0[i]<- inprod(alpha,W[i,])+gamma[1]*(betaL1[1]+b[i,1])+
      gamma[2]*(betaL2[1]+b[i,3])+
      gamma[3]*(betaL3[1]+b[i,5])+
      gamma[4]*(betaL4[1]+b[i,7])+
      gamma[5]*(betaL5[1]+b[i,9])
    
    A1[i]<- gamma[1]*(betaL1[2]+b[i,2])+
      gamma[2]*(betaL2[2]+b[i,4])+
      gamma[3]*(betaL3[2]+b[i,6])+
      gamma[4]*(betaL4[2]+b[i,8])+
      gamma[5]*(betaL5[2]+b[i,10])
    haz[i]<- nu*pow(Time[i],nu-1)*exp(A0[i]+A1[i]*Time[i]) 
    #Cumulative hazard function 
    for(j in 1:K){
      # Scaling Gauss-Kronrod/Legendre quadrature 
      xk1[i,j]<-(xk[j]+1)/2*Time[i] 
      wk1[i,j]<- wk[j]*Time[i]/2
      #  Hazard function at Gauss-Kronrod/Legendre nodes
      chaz[i,j]<- nu*pow(xk1[i,j],nu-1)*exp(A0[i]+A1[i]*xk1[i,j])
    }
    #Log-survival function with Gauss-Kronrod/Legendre requadrature
    logSurv[i]<- -inprod(wk1[i,],chaz[i,])
    #Definition of the survival log-likelihood using zeros trick
    phi[i]<-1000-death[i]*log(haz[i])-logSurv[i]
    zeros[i]~dpois(phi[i])
    #Random effects
    b[i,1:Nb]~dmnorm(mub[],Omega[,])
  }
  #Prior distributions
  for(l in 1:NbetaL){
    betaL1[l]~dnorm(0,0.001)
    betaL2[l]~dnorm(0,0.001)
    betaL3[l]~dnorm(0,0.001)
    betaL4[l]~dnorm(0,0.001)
    betaL5[l]~dnorm(0,0.001)
  }
  
  for(l in 1:nmark){ 
    gamma[l]~dnorm(0,0.001)
    tau[l]~dgamma(0.01,0.01)
    sigma[l]<-1/tau[l]
  }  
  for(l in 1:Nalpha){
    alpha[l]~dnorm(0,0.001)
  }
  Omega[1:Nb,1:Nb]~dwish(V[,],Nb)
  Sigma[1:Nb,1:Nb]<-inverse(Omega[,])
  nu~dgamma(0.01,0.01)
}
\end{lstlisting}
For running the code using jags we have used the commands given in listing \ref{hhh0}. This code includes four main parts. The data set required for running the code, the initial values for unknown parameters, the names of the parameters in the model to be monitored, and the main function "jags" serve as an interface for running JAGS analyses via the \texttt{R2jags} package. The user provides a model file, data, initial values (optional), and parameters for saving. The function requires specifying the number of Markov chains to run (n.chains), the total number of iterations per chain including burn-in (n.iter), the number of iterations at the beginning of the chain to discard (i.e., the burn-in), and the thinning rate (n.thin). We consider n.iter = 20000 iterations, with the other arguments set to their default values, where half of them are designated as burn-in and the remaining sample is thinned at a rate of 0.001 (i.e., n.thin=10). The default number of chains, which is three, is also taken into consideration.

\begin{figure}
\centering
\includegraphics[width=12cm]{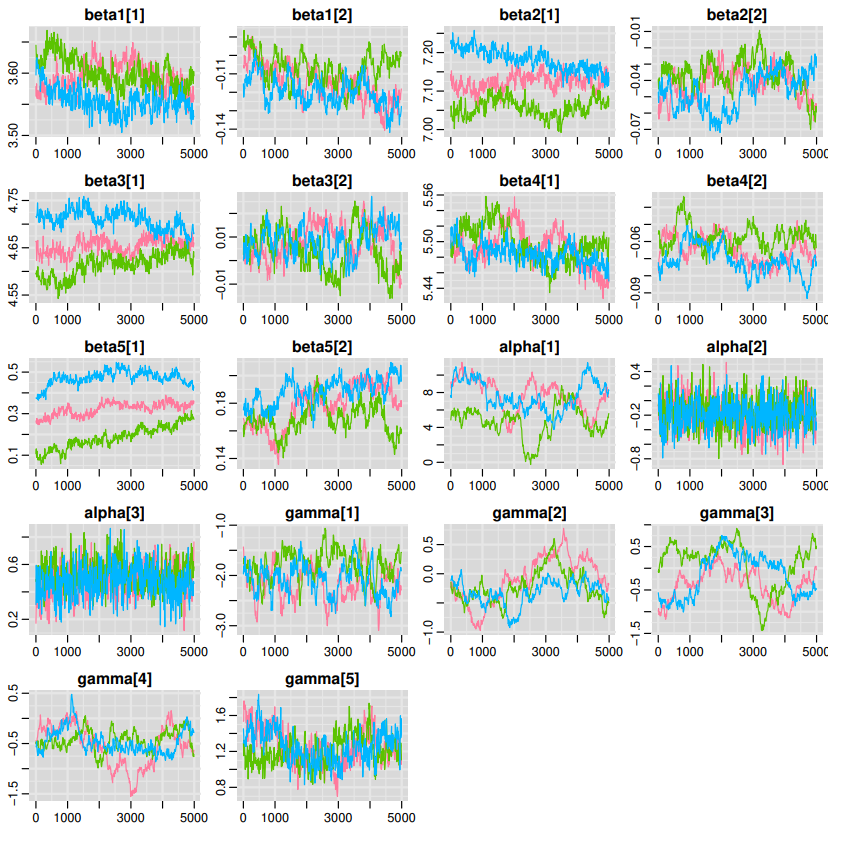}
\vspace*{-.1cm} \caption{\label{tr1}  
Trace plots from 3 chains of the 10,000 posterior samples with a thinning rate of 10 for the regression coefficients and association parameters in the PCB2 dataset. In each chain, the last 5000 samples were included.
}
\end{figure}

\begin{figure}
\centering
\includegraphics[width=12cm]{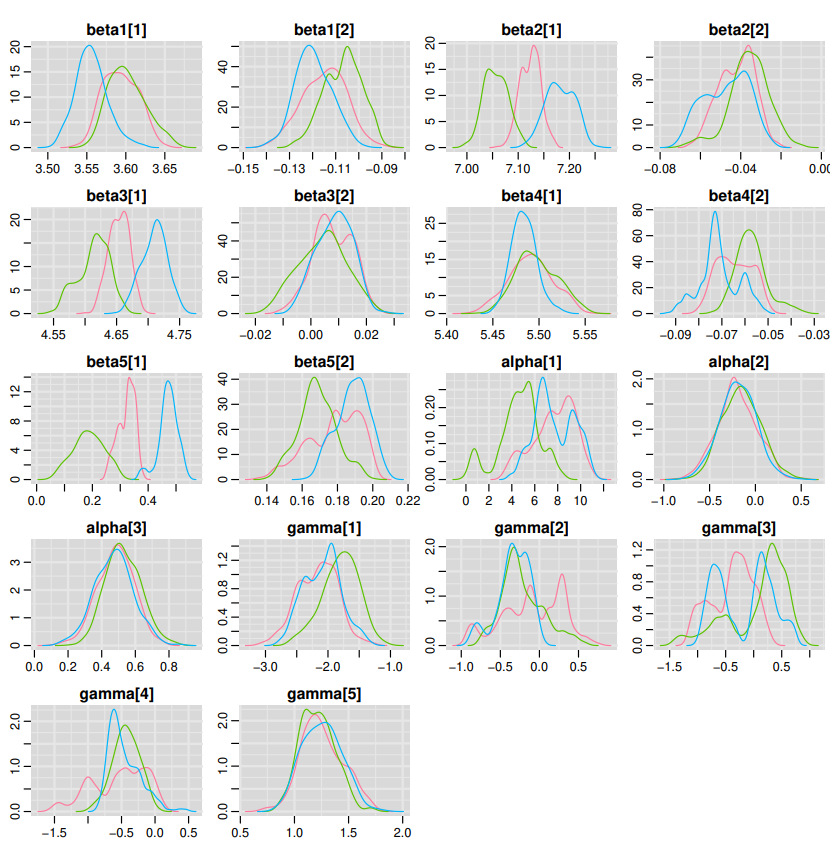}
\vspace*{-.1cm} \caption{\label{dens1}  
Density plots from 3 chains of the 10,000 posterior samples with a thinning rate of 10 for the regression coefficients and association parameters in the PCB2 dataset. In each chain, the last 5000 samples were included. }
\end{figure}

\begin{lstlisting}[caption=The R code for Bayesian implementation in  JAGS using \texttt{R2jags} package.,label=hhh0]
# Run "JAGS" from R using R2jags package
# data as list 
d.jags <- list(
  n = n, M = M, Time = Time, Y1 = Albumin, Y2 = Alkaline,
  Y3 = SGOT1, Y4 = Platelets, Y5 = SerBilir, nmark = nmark,
  XL1 = XL, ZL1 = ZL, W = W,
  death = death, mub = rep(0, 10), V = diag(1, 10), Nb = 10, zeros = rep(0, n),
  NbetaL = dim(XL)[3], Nalpha = dim(W)[2], xk = xk, wk = wk, K = K
)
# Initial value 
i.jags <- function() {
  list(
    gamma = rnorm(nmark),
    betaL1 = rnorm(dim(XL)[3]),
    betaL2 = rnorm(dim(XL)[3]),
    betaL3 = rnorm(dim(XL)[3]),
    betaL4 = rnorm(dim(XL)[3]),
    betaL5 = rnorm(dim(XL)[3]),
    alpha = rnorm(dim(W)[2]), tau = rep(1, 5), Omega = diag(runif(10))
  )
}
# Parameters of interest
parameters <- c(
  "betaL1", "betaL2", "betaL3", "betaL4", "betaL5", "alpha",
  "gamma", "nu", "sigma", "Sigma"
)
# The main fumction for rung BUGS code
jm <- jags(
  data = d.jags, inits = i.jags, parameters,
  n.iter = 20000, model.file = "model_file"
)
\end{lstlisting}
Here, we provide a detailed pipeline for conducting Bayesian inference, as well as for checking convergence and validating the results, for PBC2 data.\\
Convergence of the MCMC algorithm refers to whether the algorithm has reached its target distribution. Monitoring the convergence of the algorithm is essential for obtaining accurate results from the posterior distribution of interest. There are several methods for monitoring convergence.  There are also several R packages available for checking the convergence of MCMC output, including \texttt{coda} \cite{plummer2006coda}, \texttt{boa} \cite{smith2007boa}, \texttt{superdiag} \cite{bao2022package}, \texttt{mcmcplots} \cite{curtis2018package}, and numerous other packages.
More diagnostics are available when the model output is converted into an MCMC object. 
For converting the results to MCMC output, it is recommended to run the following command from the \texttt{coda} package.
\begin{lstlisting}[language=R] 
mcmc_output=as.mcmc(jm)
\end{lstlisting}
For cheking the convergence the first available way is to monitor the trace plots, which show the iterations plotted against the generated values.  
The convergence of the MCMC can be assumed if the values are within a zone without strong periodicities and trends  \cite{ntzoufras2011bayesian}. The following command can be used for this aim:
\begin{lstlisting}[language=R] 
traplot(mcmc_output, parms = c("beta1","beta2","beta3","beta4","beta5","alpha"))
traplot(mcmc_output, parms = c("sigma"))
traplot(mcmc_output, parms = c("nu"))
traplot(mcmc_output, parms = c("gamma"))
#traplot(mcmc_output, parms = c("Sigma"))
\end{lstlisting}
Figure \ref{tr1} shows trace plots for the regression coefficients and association parameters of three parallel chains with 10,000 iterations, where the first half is considered as burn-in. In this figure note that (alpha[1],alpha[2],alpha[3])=($\alpha_0,\alpha_1,\alpha_2$).
This figure shows that none of the regression coefficients for the longitudinal markers reach their target distribution, but $\alpha_2$ may have reached convergence. For further investigation, the posterior density plots in Figure \ref{dens1} can be helpful. These plots, also known as density plots, contain the same information despite some differences in the three chains. The density plots can be obtained using the following commands.
\begin{lstlisting}[language=R] 
denplot(mcmc_output, parms = c("beta1","beta2","beta3","beta4","beta5","alpha"))
\end{lstlisting}

\FloatBarrier

\begin{figure}
\centering
\includegraphics[width=12cm]{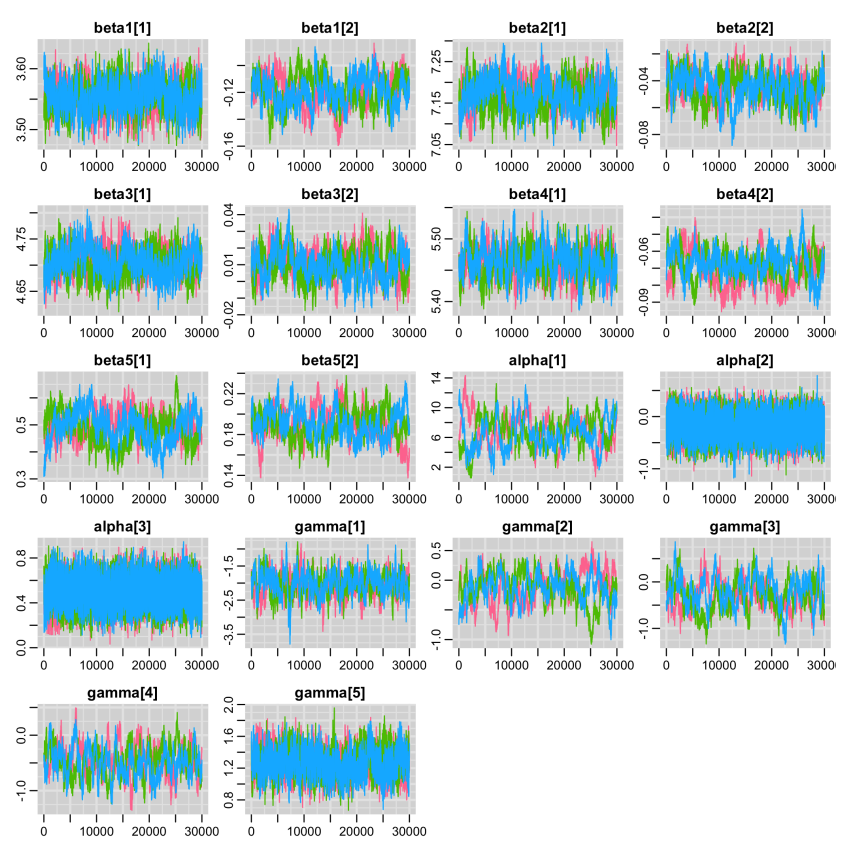}
\vspace*{-.1cm} \caption{\label{tr2}  
Trace plots from 3 chains of the 130,000 posterior samples with a thinning rate of 10 for the regression coefficients and association parameters in the PCB2 dataset. In each chain, the last 30,000 samples were included.}
\end{figure}

\begin{figure}
\centering
\includegraphics[width=12cm]{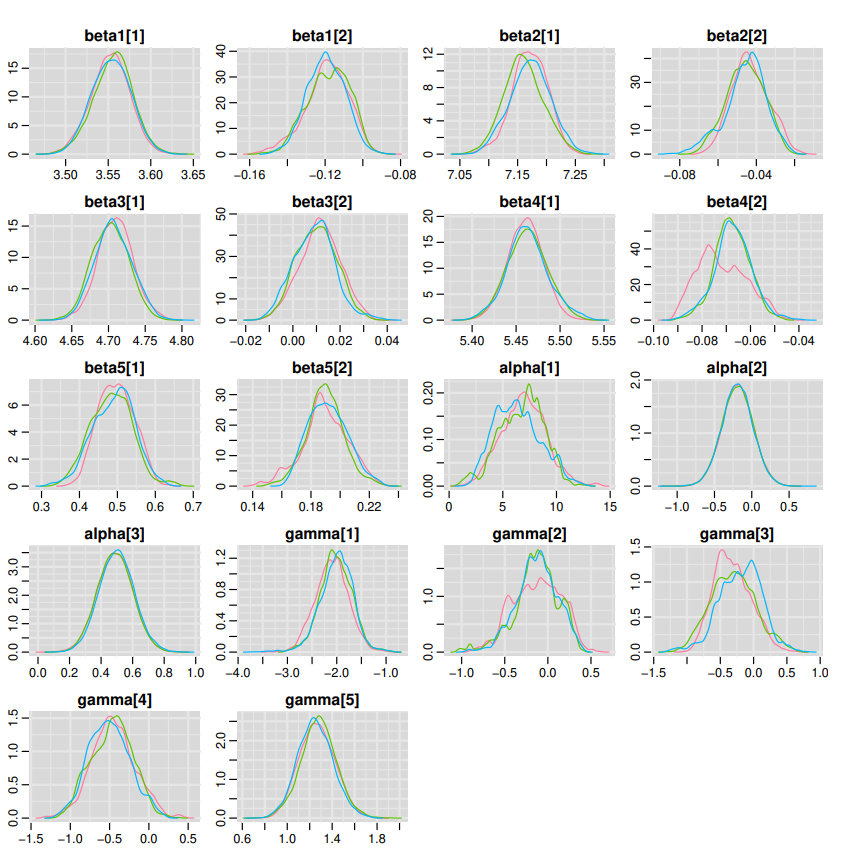}
\vspace*{-.1cm} \caption{\label{gr}  
Density plots from 3 chains of the 130,000 posterior samples with a thinning rate of 10 for the regression coefficients and association parameters in the PCB2 dataset. In each chain, the last 30,000 samples were included.}
\end{figure}

\begin{figure}
\centering
\includegraphics[width=8cm]{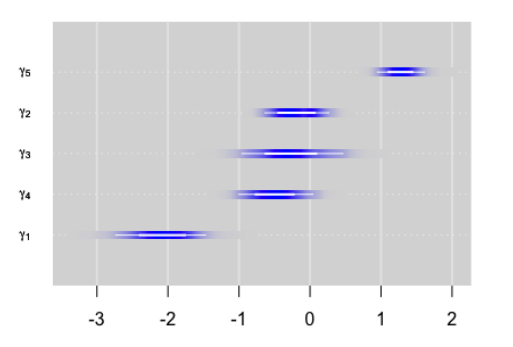}
\vspace*{-.1cm} \caption{\label{cat1}  
Caterpillar plot for association parameters $\gamma_1,\cdots,\gamma_5$. }
\end{figure}

\FloatBarrier
$~$\\
While visual tools can offer valuable insights, it is crucial to consider diagnostic tests for assessing convergence. These tests include the Gelman-Rubin  \cite{gelman1992inference}, Geweke  \cite{geweke1992evaluating}, Heidelberger-Welch  \cite{heidelberger1983simulation}, and Raftery-Lewis tests  \cite{raftery1992many}. The Gelman-Rubin diagnosis is highlighted as the primary and easily accessible criterion within the JAGS function utilized in this paper. For further details on the other tests, refer to the provided references.
The Gelman-Rubin diagnostic test compares the between-chain and within-chain variances using a test statistics called "R hat".
For the purpose of investigating convergence using this statistics, it is necessary to have two or more parallel chains with different initial values. The values of R hat for each parameter can be printed using the command "print(jm)". As chains converge to the target distributions, the R hat value is close to 1. Conversely, values larger than 1 indicate non-convergence. In fact, the Gelman-Rubin diagnosis provides the scale reduction factors for each parameter. A factor of 1 means that the between-chain variance and within-chain variance are equal, while larger values indicate a notable difference between chains. It is often said that everything below 1.1 or so is acceptable, but it's important to note that this is more of a general guideline.
We have considered using a longer chain with 130,000 iterations, with 100,000 as the burn-in period, to achieve convergence (see Figures \ref{tr2} and \ref{gr}). Also,  the \texttt{gelman.plot(.)}  function with the MCMC output as input can be used to visualize R hat values. 
After checking and confirming the convergence of MCMC samples, we can report the estimated parameters of the model, which are presented in Table \ref{estim}. This table presents posterior means, standard deviations, 95\% credible intervals, and Gelman-Rubin statistics ($\hat{R}$) for analysis. Notably, current values of Albumin and Serum bilirubin emerge as significant predictors of death. The estimated coefficient for albumin, $\hat{\gamma}_1$, stands at -2.087 with a credible interval of (-2.754, -1.459), suggesting a lower risk of death for patients with higher albumin levels. Similarly, the coefficient for serum bilirubin, $\hat{\gamma}_5$, is 1.283 with a credible interval of (0.171, 0.955), indicating elevated risks of death for patients with higher serum bilirubin values.
This inference aligns with the Caterpillar plot, provided by the command $\texttt{caterplot(mcmc\_output, parms = c("gamma"))}$ from the \texttt{mcmcplots} package, illustrated in Figure \ref{cat1}. The Caterpillar plot displays 95\% credible intervals for multiple parameters in a side-by-side bar format.
Moreover, consistent with expectations, we observed that older age contributes to an increased risk of death ($\hat{\alpha}_2$ = 0.485, credible interval: 0.262 to 0.708). However, the association between treatment and death appears weak ($\hat{\alpha}_1$ = -0.195, credible interval: -0.610 to 0.217). 
\FloatBarrier
\begin{table}[hbt!]
\centering
\footnotesize
  \caption{\label{estim} Results of the multi-marker joint model on PBC2 data: Parameter estimates (Est.), Standard deviation (SD.), lower bound of 95\% credible
interval ($2.5\%$), upper bound of 95\% credible interval ($97.5\%$), and Gelman-Rubin statistics ($\hat{R}$).}
  \begin{tabular}{cccccccc}
\hline 
&  &  & \multicolumn{5}{c}{Posterior statistics} \\ 
\cline{4-8}
&  & \multicolumn{1}{l}{Parameter} & Est. & SD. & $2.5\%$ & $97.5\%$ & $\widehat{R}$ \\ 
\hline
  \multicolumn{4}{l}{\textit{Longitudinal model for Albumin}} & & \\ 
 & \multicolumn{1}{l}{Intercept} &  $\beta_{01}$ & 3.554 & 0.023 & 3.508 & 3.598 & 1.005 \\ 
 & \multicolumn{1}{l}{Slope} & $\beta_{11}$ & -0.119 & 0.011 & -0.142 & -0.099 & 1.006 \\ 
  & \multicolumn{1}{l}{Residual variance} &   $\sigma_1^2$ & 0.099 & 0.004 & 0.092 & 0.107 & 1.001 \\ 
    \\
  \multicolumn{6}{l}{\textit{Longitudinal model for Alkaline}} \\ 
  & \multicolumn{1}{l}{Intercept} & $\beta_{02}$ & 7.168 & 0.034 & 7.099 & 7.234 & 1.028 \\ 
 & \multicolumn{1}{l}{Slope} &  $\beta_{12}$ & -0.045 & 0.010 & -0.066 & -0.026 & 1.015 \\ 
  & \multicolumn{1}{l}{Residual variance} & $\sigma_2^2$ & 0.104 & 0.004 & 0.097 & 0.113 & 1.002 \\ 
\\
    \multicolumn{6}{l}{\textit{Longitudinal model for SGOT}} \\ 
 & \multicolumn{1}{l}{Intercept} &  $\beta_{03}$ & 4.706 & 0.025 & 4.658 & 4.756 & 1.015 \\ 
  & \multicolumn{1}{l}{Slope} & $\beta_{13}$ & 0.010 & 0.009 & -0.007 & 0.028 & 1.015 \\ 
  & \multicolumn{1}{l}{Residual variance} &  $\sigma_3^2$ & 0.073 & 0.003 & 0.068 & 0.079 & 1.001 \\ 
\\
    \multicolumn{6}{l}{\textit{Longitudinal model for Platelets}} \\ 
  & \multicolumn{1}{l}{Intercept} & $\beta_{04}$ &  5.462 & 0.022 & 5.418 & 5.508 & 1.007 \\ 
 & \multicolumn{1}{l}{Slope} &  $\beta_{14}$ & -0.069 & 0.009 & -0.086 & -0.052 & 1.085 \\ 
  & \multicolumn{1}{l}{Residual variance} & $\sigma_4^2$ & 0.042 & 0.002 & 0.039 & 0.045 & 1.001 \\ 
\\
    \multicolumn{6}{l}{\textit{Longitudinal model for Serum bilirubin}} \\ 
 & \multicolumn{1}{l}{Intercept} &  $\beta_{05}$ & 0.491 & 0.053 & 0.386 & 0.590 & 1.017 \\ 
  & \multicolumn{1}{l}{Slope} & $\beta_{15}$ & 0.190 & 0.014 & 0.162 & 0.219 & 1.015 \\ 
  & \multicolumn{1}{l}{Residual variance} &   $\sigma_5^2$ & 0.119 & 0.005 & 0.111 & 0.128 & 1.001 \\ 
\\
      \multicolumn{6}{l}{\textit{Survival model}} \\ 
 & \multicolumn{1}{l}{Intercept}  & $\alpha_0$ & 6.695 & 2.108 & 2.700 & 10.749 & 1.014 \\ 
 & \multicolumn{1}{l}{Drug} & $\alpha_1$ & -0.192 & 0.208 & -0.601 & 0.217 & 1.001 \\ 
 & \multicolumn{1}{l}{Age} & $\alpha_2$ & 0.493 & 0.111 & 0.276 & 0.712 & 1.003 \\ 
 & \multicolumn{1}{l}{Current value of Albumin} &  $\gamma_1$ & -2.040 & 0.336 & -2.717 & -1.412 & 1.017 \\ 
 & \multicolumn{1}{l}{Current value of Alkaline} & $\gamma_2$ &  -0.130 & 0.256 & -0.671 & 0.314 & 1.004 \\ 
  & \multicolumn{1}{l}{Current value of SGOT} & $\gamma_3$ & -0.274 & 0.318 & -0.880 & 0.359 & 1.036 \\ 
  & \multicolumn{1}{l}{Current value of Platelets} & $\gamma_4$ & -0.475 & 0.269 & -0.991 & 0.055 & 1.008 \\ 
 & \multicolumn{1}{l}{Current value of Serum bilirubin} &  $\gamma_5$ & 1.266 & 0.158 & 0.961 & 1.585 & 1.007 \\ 
  & \multicolumn{1}{l}{Shape} &   $\nu$ & 0.837 & 0.092 & 0.664 & 1.023 & 1.002 \\ 
   \hline
     \end{tabular}
\end{table}
\FloatBarrier

\section{JM for competing risks outcome}
\subsection{Model specification}
In this section, we consider joint modeling for a Gaussian longitudinal outcome and competing risks with $L$ different failure causes. However, the model and the BUGS syntax can be easily extended to handle situations with multivariate longitudinal outcomes by referring to  Section 3.\\
Let $T_{i1}^*,\ldots,T_{iL}^*$ represent the true failure times for $L$ different causes of failure for subject $i$.
The observed time is $T_i=min(T_{i1}^*,\ldots,T_{iL}^*,C_i)$ where $C_i$ denotes the censoring time. Also, the event indicator is denoted as $\delta_i\in \{0,1,\ldots,L\}$, where $0$ corresponds to censoring and $1$ to $L$ denotes each cause of the competing risks. Denote the longitudinal marker for the $i$th subject at time $s_{ij}$ by $Y_{ij}={Y}_{i} (s_{ij})$. As a result, consider $\{T_i,\delta_i,{y}_{i},i=1,~\ldots,n\}$ a sample from the target population.\\
We assume the linear mixed-effects model of equation \eqref{long} for the longitudinal outcome. For each of the $L$ causes, the following  cause-specific proportional hazard model is postulated:
\begin{eqnarray}\label{csh}
\lambda_{il}(t|{w}_i,{b}_i)=\lambda_{0l}(t)\exp\left({\alpha}^{\top}_l{w}_i+g\left(\mathcal{H}_i(t), {b}_i, {\gamma}_l\right )\right), \quad t>0,
\end{eqnarray}
where $\lambda_{0l}(t)$ represents the baseline hazard function for the $l$th cause, ${w}_i$ is a vector of exogenous covariates with corresponding regression coefficients ${\alpha}_l$, $g(.)$ is a known function of $\mathcal{H}_i(t)$ and the random effects ${b}_{i}$ depending on parameters ${\gamma}_l$, which defines the dependence structure between cause $l$ and the marker. The prior distributions for the parameters of longitudinal model are the same as those discussed in Section \ref{s11}. For the cause-specific hazard model, we assume that ${\alpha}_l\sim \mathcal{N} ({{\mu}_{\alpha}}_l,{{\Sigma}_{\alpha}}_l)$, ${\gamma}_l\sim \mathcal{N} ({{\mu}_{\gamma}}_l,{{\Sigma}_{\gamma}}_l)$, and the vector of parameters for the baseline hazard of cause $l$, ${\theta}_{\lambda_{0l}}\sim \pi({\theta}_{\lambda_{0l}})$. The joint posterior distribution of ${\theta}=({\beta},\sigma^2,{\Sigma},{\alpha}_l,{\gamma}_l,{\theta}_{\lambda_{0l}},l=1,\ldots,L)$ for the joint modeling of the Gaussian longitudinal model \eqref{long} and the cause-specific hazard model \eqref{csh} is given by:
\begin{eqnarray}\label{postcsh}
\begin{aligned}
& 
\pi({\theta}, {b} \mid {y}, {t}, {\delta}) \\ & \propto \prod_{i=1}^n\left( \prod_{j=1}^{n_i} \phi\left(y_{i j };\mu_i(s_{ij}), \sigma^2\right) \right) \\
& \times \left( \prod_{l=1}^{L} \lambda_{il}\left(t_i \mid {w}_i, {b}_i\right)^{{\rm{I}}(\delta_i=l)} S(t_i \mid {w}_i, {b}_i)\right)  \\
& \times \phi\left({b}_i ; \mathbf{0}, {D}\right) \times \pi({\theta}),\\
& 
\end{aligned}
\end{eqnarray}
where $S(t_i \mid {w}_i, {b}_i)=\exp \left(-\sum_{l=1}^{L} \Lambda_{il}\left(t_i \mid {w}_i, {b}_i\right)\right)$.

\subsection{Codes for model estimation}
To simulate the data, we define a scenario with 
$n=500$ subjects and $L=2$ causes. The longitudinal outcomes follow the model represented by equation \eqref{gau}. For the cause-specific survival outcomes, we employ the model:
\begin{eqnarray}\label{csh1}
\lambda_{il}(t)=\lambda_{0l}\exp\left(\gamma_l \mu_i(t)\right), \quad t>0,~~l=1,2.
\end{eqnarray}
In generating data for each cause, we utilize the inverse of the cumulative hazard function, elaborated upon in Section \ref{simuni}. The final survival time is determined as the minimum of the two generated values, along with administrative censoring.
 The real values and the R code for generating data can be found in 
\url{https://github.com/tbaghfalaki/JM-with-BUGS-and-JAGS/tree/main/CR}.\\
For data analysis, a cause-specific proportional hazard sub-model is considered. Although different forms of baseline hazards can be considered, here we will focus on two: a constant baseline hazard and a Weibull baseline hazard. At first, we consider the cause-specific proportional hazard sub-model with constant baseline hazards. A zero-trick is used for this purpose. The BUGS code for the longitudinal sub-model is the same as that in Section \ref{simuni}, the values of A0l[i], A1l[i], l=1,2, and the mean of the Poisson outcome for the zero-trick are given in Listing \ref{cs1}. Also, the complete R code is available at \url{https://github.com/tbaghfalaki/JM-with-BUGS-and-JAGS/tree/main/CR}.
\begin{lstlisting}[caption=The BUGS code for the  cause-specific hazard sub-model with constant baseline hazards.,label=cs1]
   #Survival and censoring times
    # 1th cause
    A01[i]<- inprod(alpha1[],W[i,])+gamma1*(betaL1[1]+betaL1[3]*x1[i]+betaL1[4]*x2[i]+b[i,1])
    A11[i]<- gamma1*(betaL1[2]+b[i,2])
    haz1[i]<- exp(A01[i]+A11[i]*Time[i])
    chaz1[i]<- exp(A01[i])/A11[i]*(exp(A11[i]*Time[i])-1)
    logSurv1[i]<- -chaz1[i]
    
    # 2th cause
    A02[i]<- inprod(alpha2[],W[i,])+gamma2*(betaL1[1]+betaL1[3]*x1[i]+betaL1[4]*x2[i]+b[i,1])
    A12[i]<- gamma2*(betaL1[2]+b[i,2])
    haz2[i]<- exp(A02[i]+A12[i]*Time[i])
    chaz2[i]<- exp(A02[i])/A12[i]*(exp(A12[i]*Time[i])-1)

    logSurv2[i]<- -chaz2[i]   
    #Definition of the survival log-likelihood using zeros trick
    phi[i]<-100000-equals(CR[i],1)*log(haz1[i])-equals(CR[i],2)*log(haz2[i])-logSurv1[i]-logSurv2[i]
\end{lstlisting}
For the cause-specific hazard model with a Weibull baseline hazard, in addition to the zero-trick, a Gaussian quadrature should be used to approximate the integral for computing the cumulative hazard. The BUGS code for this section is provided in Listing \ref{cs2}. 
Also, the complete R code is available at \url{https://github.com/tbaghfalaki/JM-with-BUGS-and-JAGS/tree/main/CR}.
\begin{lstlisting}[caption=The BUGS code for the  cause-specific hazard sub-model with Weibull baseline hazards.,label=cs2]
#Survival and censoring times
    # 1th cause
    A01[i]<- inprod(alpha1[],W[i,])+gamma1*(betaL1[1]+betaL1[3]*x1[i]+betaL1[4]*x2[i]+b[i,1])
    A11[i]<- gamma1*(betaL1[2]+b[i,2])
    haz1[i]<- nu1*pow(Time[i],nu1-1)*exp(A01[i]+A11[i]*Time[i])
    for(j in 1:K){
      xk1[i,j]<-(xk[j]+1)/2*Time[i] 
      wk1[i,j]<- wk[j]*Time[i]/2
      chaz1[i,j]<- nu1*pow(xk1[i,j],nu1-1)*exp(A01[i]+A11[i]*xk1[i,j])
      chaz2[i,j]<- nu2*pow(xk1[i,j],nu2-1)*exp(A02[i]+A12[i]*xk1[i,j])
    }
    logSurv1[i]<- -inprod(wk1[i,],chaz1[i,])
    # 2th cause
    A02[i]<- inprod(alpha2[],W[i,])+gamma2*(betaL1[1]+betaL1[3]*x1[i]+betaL1[4]*x2[i]+b[i,1])
    A12[i]<- gamma2*(betaL1[2]+b[i,2])
    haz2[i]<- nu2*pow(Time[i],nu2-1)*exp(A02[i]+A12[i]*Time[i])
    logSurv2[i]<- -inprod(wk1[i,],chaz2[i,])    
    #Definition of the survival log-likelihood using zeros trick
    phi[i]<-100000-equals(CR[i],1)*log(haz1[i])-equals(CR[i],2)*log(haz2[i])-logSurv1[i]-logSurv2[i]  
\end{lstlisting}
\FloatBarrier

\subsection{Application to PBC2 data}
\begin{figure}
\centering
\includegraphics[width=15cm]{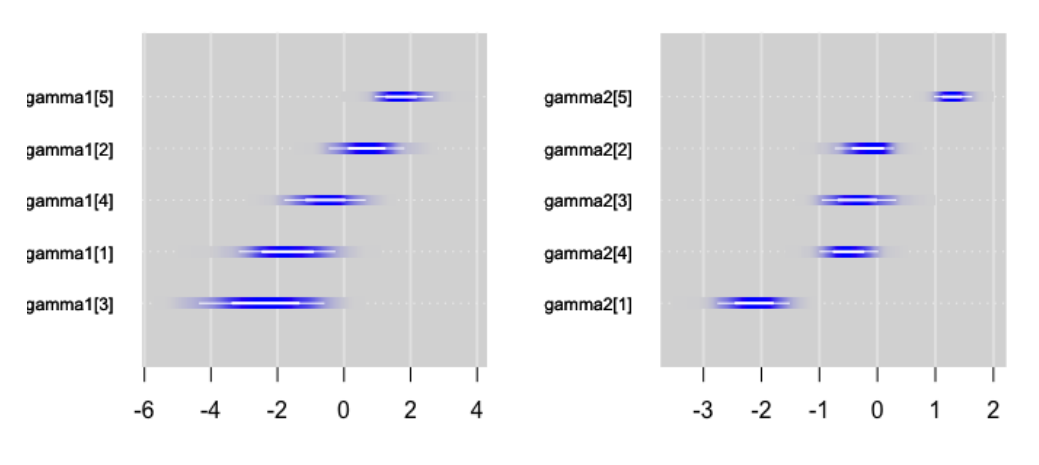}
\vspace*{-.1cm} \caption{\label{cat2}  
Caterpillar plot illustrating association parameters for transplanted patients (left panel) and deceased patients (right panel).}
\end{figure}

Here, we revisit the PBC2 data set described in Section \ref{pbc2_sec}. We considered the five longitudinal markers described in that section. We also considered the same model as before for them. For the time-to-event data, instead of focusing solely on time to death, two outcomes are considered: death and transplantation. Out of 312 patients, 140 died and 29 received transplants. A cause-specific hazard model was used, in which the hazard depends on the current values of the included markers and is adjusted for the drug and the patient's standardized age at enrollment. \\
For the Bayesian analysis, a total of 70,000 iterations were conducted, with the initial 35,000 iterations considered as the burn-in period to stabilize the results. The outcomes are presented in Table \ref{estimcr}, while Figure \ref{cat2} illustrates the Caterpillar plots of the association parameters.
The analysis reveals significant factors influencing both the risk of transplantation and death. Notably, albumin, SGOT, and serum bilirubin stand out as influential factors for transplantation. Higher levels of albumin and SGOT, along with lower levels of serum bilirubin, correspond to reduced risks of transplantation. Similarly, albumin and serum bilirubin emerge as predictors of death. Higher levels of albumin and lower levels of serum bilirubin are associated with decreased risks of death.
Furthermore, in alignment with the conclusions drawn in Section \ref{pbc2_sec}, treatment is poorly associated with both events. Conversely, older age is associated with a lower risk of transplantation and a higher risk of death.
The complete R code of this application is available at \url{https://github.com/tbaghfalaki/JM-with-BUGS-and-JAGS/blob/main/CR/data_preparation_and_jags.R}.

\begin{table}[hbt!]
\centering
\footnotesize
  \caption{\label{estimcr} Results of the multi-marker joint model considering competing risks on PBC2 data: Parameter estimates (Est.), Standard deviation (SD.), lower bound of the 95\% credible interval (2.5\%), upper bound of the 95\% credible interval (97.5\%), and Gelman-Rubin statistics ($\hat{R}$).}
  \begin{tabular}{cccccccc}
\hline 
&  &  & \multicolumn{5}{c}{Posterior statistics} \\ 
\cline{4-8}
&  & \multicolumn{1}{l}{Parameter} & Est. & SD. & $2.5\%$ & $97.5\%$ & $\widehat{R}$ \\ 
\hline
  \multicolumn{4}{l}{\textit{Longitudinal model for Albumin}} & & \\ 
 & \multicolumn{1}{l}{Intercept} & $\beta_{01}$  & 3.559 & 0.021 & 3.518 & 3.602 & 1.003 \\ 
 & \multicolumn{1}{l}{Slope} &  $\beta_{11}$  & -0.123 & 0.009 & -0.140 & -0.105 & 1.009 \\ 
  & \multicolumn{1}{l}{Residual variance} &  $\sigma_1^2$ & 0.099 & 0.004 & 0.092 & 0.106 & 1.001 \\ 
  \\
  \multicolumn{6}{l}{\textit{Longitudinal model for Alkaline}} \\ 
  & \multicolumn{1}{l}{Intercept} & $\beta_{02}$  & 7.167 & 0.035 & 7.093 & 7.232 & 1.015 \\ 
  & \multicolumn{1}{l}{Slope} & $\beta_{12}$  & -0.044 & 0.010 & -0.063 & -0.023 & 1.031 \\
    & \multicolumn{1}{l}{Residual variance} &  $\sigma_2^2$ & 0.104 & 0.004 & 0.097 & 0.112 & 1.007 \\ 
\\
  \multicolumn{6}{l}{\textit{Longitudinal model for SGOT}} \\ 
 & \multicolumn{1}{l}{Intercept} &   $\beta_{03}$  & 4.705 & 0.027 & 4.652 & 4.760 & 1.022 \\ 
 & \multicolumn{1}{l}{Slope} &  $\beta_{13}$  & 0.008 & 0.009 & -0.010 & 0.026 & 1.006 \\ 
    & \multicolumn{1}{l}{Residual variance} &  $\sigma_3^2$ & 0.073 & 0.003 & 0.068 & 0.079 & 1.001 \\ 
\\
    \multicolumn{6}{l}{\textit{Longitudinal model for Platelets}} \\ 
  & \multicolumn{1}{l}{Intercept} & $\beta_{04}$  & 5.465 & 0.022 & 5.420 & 5.509 & 1.032 \\ 
 & \multicolumn{1}{l}{Slope} &  $\beta_{14}$  & -0.071 & 0.008 & -0.085 & -0.055 & 1.001 \\ 
    & \multicolumn{1}{l}{Residual variance} &  $\sigma_4^2$ & 0.042 & 0.002 & 0.039 & 0.045 & 1.002 \\ 
\\
    \multicolumn{6}{l}{\textit{Longitudinal model for Serum bilirubin}} \\ 
 & \multicolumn{1}{l}{Intercept} &  $\beta_{05}$  & 0.478 & 0.055 & 0.368 & 0.581 & 1.011 \\ 
 & \multicolumn{1}{l}{Slope} &  $\beta_{15}$  & 0.192 & 0.013 & 0.167 & 0.215 & 1.069 \\ 
  & \multicolumn{1}{l}{Residual variance} &  $\sigma_5^2$ & 0.119 & 0.004 & 0.111 & 0.128 & 1.001 \\
  \\
    \multicolumn{6}{l}{\textit{Cause-specific model for transplant}} \\ 
 & \multicolumn{1}{l}{Intercept}  & $\alpha_{10}$ & 6.389 & 5.016 & -3.592 & 15.619 & 1.007 \\ 
  & \multicolumn{1}{l}{Drug}  &    $\alpha_{11}$ & -0.172 & 0.440 & -1.052 & 0.692 & 1.001 \\ 
 & \multicolumn{1}{l}{Age} &    $\alpha_{12}$ & -1.326 & 0.336 & -2.032 & -0.726 & 1.002 \\ 
 & \multicolumn{1}{l}{Current value of Albumin} &  $\gamma_{11}$ & -1.706 & 0.747 & -3.141 & -0.280 & 1.020 \\ 
 & \multicolumn{1}{l}{Current value of Alkaline} & $\gamma_{12}$ & 0.686 & 0.548 & -0.433 & 1.795 & 1.054 \\ 
  & \multicolumn{1}{l}{Current value of SGOT} & $\gamma_{13}$ & -2.369 & 0.957 & -4.341 & -0.608 & 1.015 \\ 
 & \multicolumn{1}{l}{Current value of Platelets} & $\gamma_{14}$ & -0.548 & 0.597 & -1.777 & 0.628 & 1.019 \\ 
 & \multicolumn{1}{l}{Current value of Serum bilirubin} & $\gamma_{15}$ & 1.713 & 0.442 & 0.945 & 2.650 & 1.010 \\  
    & \multicolumn{1}{l}{Shape} &  $\nu_1$ & 1.246 & 0.310 & 0.674 & 1.863 & 1.014 \\ 
\\
    \multicolumn{6}{l}{\textit{Cause-specific model for death}} \\ 
 & \multicolumn{1}{l}{Intercept}  &  $\alpha_{20}$ & 7.644 & 1.786 & 4.338 & 11.164 & 1.010 \\ 
 & \multicolumn{1}{l}{Drug}  &   $\alpha_{21}$ & -0.159 & 0.206 & -0.573 & 0.234 & 1.000 \\ 
  & \multicolumn{1}{l}{Age} &  $\alpha_{22}$ & 0.460 & 0.109 & 0.249 & 0.667 & 1.001 \\ 
 & \multicolumn{1}{l}{Current value of Albumin} & $\gamma_{21}$ & -2.122 & 0.318 & -2.746 & -1.525 & 1.001 \\ 
 & \multicolumn{1}{l}{Current value of Alkaline} & $\gamma_{22}$ & -0.161 & 0.258 & -0.723 & 0.274 & 1.048 \\ 
  & \multicolumn{1}{l}{Current value of SGOT} & $\gamma_{23}$ & -0.346 & 0.326 & -0.949 & 0.314 & 1.013 \\ 
 & \multicolumn{1}{l}{Current value of Platelets} & $\gamma_{24}$ & -0.505 & 0.255 & -0.989 & 0.003 & 1.006 \\ 
 & \multicolumn{1}{l}{Current value of Serum bilirubin} & $\gamma_{25}$ & 1.289 & 0.162 & 0.987 & 1.623 & 1.001 \\ 
   & \multicolumn{1}{l}{Shape} &   $\nu_2$ & 0.822 & 0.087 & 0.654 & 0.995 & 1.007 \\ 
   \hline
     \end{tabular}
\end{table}
\FloatBarrier

\FloatBarrier

\section{JM of zero-inflated longitudinal marker and survival time }\label{ZIJM1}
In medical studies, particularly those examining hospital admissions, medical expenses, disease counts, or the analysis of count biomarkers, datasets often contain an abundance of zeros. These zeros can arise from diverse factors, including structural processes leading to no events and counting processes.
Traditional count models, such as Poisson or negative binomial distributions, may fail to adequately address the excess zero phenomenon, which can result in biased or inefficient estimations. Zero-inflated models offer a robust solution by explicitly modeling the excess zeros as a separate process from the count distribution. This approach captures the underlying complexities, thereby improving the accuracy of statistical inference. These models are particularly valuable in situations where the occurrence of events is rare, but the presence of excess zeros significantly impacts the overall data distribution. By incorporating zero-inflated models, researchers can better understand the factors influencing the probability of observing zeros and the mean or median of the count outcome. This leads to more reliable conclusions in medical research and biostatistical analyses.\\
In this section, we explore a joint modeling of zero-inflated longitudinal count measurements and a time-to-event outcome. Implementing joint modeling  as outlined by Baghfalaki and Ganjali \cite{baghfalaki2021approximate}, we combine a zero-inflated negative binomial model with a proportional hazard model for the time-to-event outcome. This joint approach incorporates a shared random effects model, providing a comprehensive understanding of their combined relationship. While the joint modeling  framework offers flexibility for various structural setups, our discussion will focus on this specific model.
\subsection{Model specification }
Let $Y_{i}(s_{ij})$ represent the longitudinal measurements for the $i^{\rm{th}}$ subject at time $s_{ij}$, $i=1,2,\cdots,n$. Also, the time-to-event outcome consists of the pairs $\{(T_i, \delta_i), i=1,\cdots,n\}$.\\
Let the longitudinal measurements be recorded using a count variable with excess zeros. A hurdle model is considered for modeling zero-inflated count data, which is a finite mixture:
\begin{eqnarray}\label{ziw1}
{Y_{ij}} \sim \left\{ {\begin{array}{*{20}{c}}
   0 & {{\rm{with ~probability~ }}{\pi (s_{ij})}}  \\
   {P_{TY}({Y_{ij}} = {y}|{\kappa (s_{ij})},r)} & {{\rm{with~ probability~  }}{1 -\pi  (s_{ij})}~\rm{for}~y>0}  \\
\end{array}} \right.,
\end{eqnarray}
where 
$$P_{TY}({Y_{ij}} = {y}|{\kappa (s_{ij})},r) = \frac{{\Gamma ({y} + r)}}{{{y}!\Gamma (r)(1 - \kappa (s_{ij})^r)}}\kappa (s_{ij})^r{(1 - \kappa (s_{ij}))^{{y}}},{y} = 1,2, \cdots$$
is the truncated negative binomial distribution at zero, $r>0$ and $0<\kappa (s_{ij})<1$. 
Under the assumption of a negative binomial distribution for the count outcome, $Y_{ij}$, we have $E[Y(s_{ij})]=\frac{r(1 - \kappa (s_{ij}))}{\kappa (s_{ij})}$ and $Var[Y(s_{ij})]=\eta (s_{ij})\left(\eta (s_{ij})+r\right)$, where $\eta (s_{ij}) = E[Y(s_{ij})]$ and $1+\frac{\eta (s_{ij})}{r}$ represents the dispersion parameter.
 The zero-inflated negative binomial model can be derived by assuming the following models for the mean and probability, respectively:
\begin{eqnarray}\label{zi2}
\begin{array}{l}
 \log ({\eta }({s_{ij}})) = {{x}_{1i}}({s_{ij}}){{\beta} _1} + b_{i1}, \\
 {\rm{logit}}(\pi ({s_{ij}})) = {{x}_{2i}}({s_{ij}}){{\beta} _2} + b_{i2}, \\
 \end{array}
\end{eqnarray}
We use the notation $Y_{ij}\sim ZINB(\pi(s_{ij}),\eta(s_{ij}),r)$ to denote the zero-inflated negative binomial model.\\
For the time-to-event process, a proportional hazard model is considered. The hazard function in our proposed model is given by:
\begin{eqnarray}\label{zi3}
\lambda_{i}(t|{w}_i)=\lambda_{0}(t)\exp\left({\alpha}^{\top}{w}_i+\gamma_1 b_{i1}+\gamma_2 b_{i2}\right),
\end{eqnarray}
where ${w}_i$ is a vector of covariates.  
The association among longitudinal outcomes within subjects is addressed by $b_{i1}$ and $b_{i2}$, while the relationship between the zero-inflated longitudinal measurements and survival outcomes is captured by a linear combination of these random effects, where ${b}_{i}=(b_{i1},b_{i2})\sim \mathcal{N}_2({0},{D})$. Let the parameters of the model be ${\theta}=({\beta}_1,~{\beta}_2,~{\alpha},\gamma_1,\gamma_2,{\theta}_{\lambda_0})$, the random effects for all individuals are denoted as ${b}$. For deriving the likelihood function for the hurdle model, we define an indicator function as follows:
$$\mathcal{z}_{i j} = \begin{cases}1 & Y_{i j} = 0 \\ 0 & Y_{i j} \neq 0\end{cases}.$$
Therefore, the joint posterior distribution is given by:
\begin{eqnarray}\label{post3}
\begin{aligned}
& 
\pi({\theta},  {b} \mid {y},{z}, {t}, {\delta}) \\ & \propto \prod_{i=1}^n\left( \prod_{j=1}^{n_i} {\pi (s_{ij})}^{\mathcal{z}_{ij}}\times
\left( (1-\pi(s_{ij})) \times P_{TY}({Y_{ij}} = {y_{ij}}|{\kappa (s_{ij})},r)\right)^{1-\mathcal{z}_{ij}} \right) \\
& \times \lambda_i\left(t_i \mid {w}_i, {b}_i\right)^{\delta_i} \exp \left(-\Lambda_i\left(t_i \mid {w}_i, {b}_i\right)\right) \\
& \times \phi\left({b}_i ; \mathbf{0}, {D}\right) \times \pi({\theta}).\\
& 
\end{aligned}
\end{eqnarray} 
where, we assume specific prior distributions for the parameters. The coefficients ${\beta}_1$ and ${\beta}_2$ follow multivariate normal distributions, specifically ${\beta}_1\sim \mathcal{N} ({\mu}_{\beta_1},{\Sigma}_{\beta_1})$ and ${\beta}_2\sim \mathcal{N} ({\mu}_{\beta_2},{\Sigma}_{\beta_2})$. Additionally, the survival sub-model incorporates assumptions where ${\alpha}\sim \mathcal{N} ({\mu}_{\alpha},{\Sigma}_{\alpha})$, ${\gamma}_j$ (for $j=1,2$) follows independent normal distributions, $\mathcal{N} ({\mu}_{\gamma_j},{\sigma^2}_{\gamma_j}),~j=1,2$, and the covariance matrix ${D}$ is assigned an inverse Wishart distribution, ${D}\sim \mathcal{IW}({\Omega}_{D},\omega_{D})$.

\subsection{Codes for model estimation}
For generating the data we consider the following model:
\begin{eqnarray}\label{s2}
\begin{array}{l}
 \log (\eta ({s_{ij}})) = {\beta _{10}} + {\beta _{11}}{s_{ij}} + {\beta _{12}}{x_{1i}} + {\beta _{13}}{x_{2i}} + {b_{i1}}, \\
{ \rm{logit}} (\pi ({s_{ij}})) = {\beta _{20}} + {\beta _{21}}{s_{ij}} + {\beta _{22}}{x_{1i}} +  {b_{i2}}, \\
 \end{array}
 \end{eqnarray}
and
\begin{eqnarray}\label{s4}
\lambda({t_i}) = {\lambda_0}({t_i})\exp \left\{ {{\alpha _{0}} + {\alpha _{1}}{x_{1i}} + \gamma_1 {b_{i1}} + \gamma_2 {b_{i2}}} \right\},
\end{eqnarray}
such that ${x}_{1i},~i=1,2,\cdots,n$ is a binary explanatory variable generated by ${x}_{1i} \sim Ber (0.5)$ and ${x}_{2i},~i=1,2,\cdots,n$ is generated by $\mathcal{N}(0,1)$. The maximum number of repeated measures is $9$ such that $s_{ij}\in\{0, 0.125, 0.250,\ldots, 1\}$. Also, the function $\lambda_0(t)=\nu t^{\nu-1}$ is used, which leads to the Weibull hazard function. Additionally, almost $30\%$ right censoring is considered. The real values of the parameters and the R code for generating the data can be found in \url{https://github.com/tbaghfalaki/JM-with-BUGS-and-JAGS/tree/main/Zero_inflation}.\\
We consider a zero-trick for estimating procedure. The BUGS code for this part is given in Listing \ref{hhh}. Also, the full R code is given in \url{https://github.com/tbaghfalaki/JM-with-BUGS-and-JAGS/tree/main/Zero_inflation}.
\begin{lstlisting}[caption=The BUGS code for
JM for zero-inflated longitudinal marker with a Weibull baseline hazards.,label=hhh]
model{
  K<-1000
  for (i in 1:n) {
    for (j in 1:M[i]) {
      zeros[i,j]~dpois(phi[i,j])
      phi[i,j]<-  - ll[i,j]+K				
      ll[i,j]<-z[i,j]*log(pi[i,j]) +
        (1-z[i,j])*(log(1-pi[i,j])+loggam(r+y[i,j])-loggam(r)-loggam(y[i,j]+1)+
                      r*log(r/(r+lambda[i, j]))+
                      y[i,j]*log(lambda[i, j]/(lambda[i, j]+r))-log(1-pow(r/(r+lambda[i, j]),r)))
      
      log(lambda[i, j])<-beta1[1]+beta1[2]*t[j]+beta1[3]*x1[i]+beta1[4]*x2[i]+
        U[i,1]
      logit(pi[i, j])<-beta2[1]+beta2[2]*t[j]+beta2[3]*x1[i]+U[i,2]
    }
    surt[i] ~ dweib(nu,mut[i])    
    is.censored[i]~dinterval(surt[i],c[i])
    log(mut[i])<-alpha[1]+alpha[2]*x1[i]+gamma[1]*U[i, 1]+gamma[2]*U[i, 2]
    U[i,1:2] ~ dmnorm(U0[],tau[,])   
  }  
# priors
  r~ dgamma(0.1,0.1) 
  nu ~ dgamma(0.1,0.1) 
  sigma[1:2,1:2]<-inverse(tau[,])
  tau[1:2,1:2] ~ dwish(R[,], 2)
  for(k in 1:Nbeta1){ 
    beta1[k]~dnorm(0, 0.0001)}
  for(k in 1:Nbeta2){ 
    beta2[k]~dnorm(0, 0.0001)}
  for(k in 1:Nalpha){ 
    alpha[k]~dnorm(0, 0.0001)}
  for(k in 1:Ngamma){ 
    gamma[k]~dnorm(0, 0.0001)}
}
\end{lstlisting}

\subsection{Application to pregnancy data}
\begin{figure}[hbt!]
\centering
\includegraphics[width=8cm]{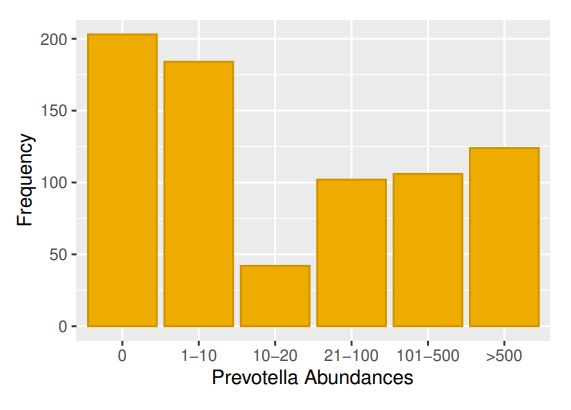}
\vspace*{-.1cm} \caption{\label{barpreg}
Bar chart of Prevotella abundances in pregnancy data where for better visualization, data larger than zero are converted into intervals. }
\end{figure}

\begin{figure}[hbt!]
\centering
\includegraphics[width=9cm]{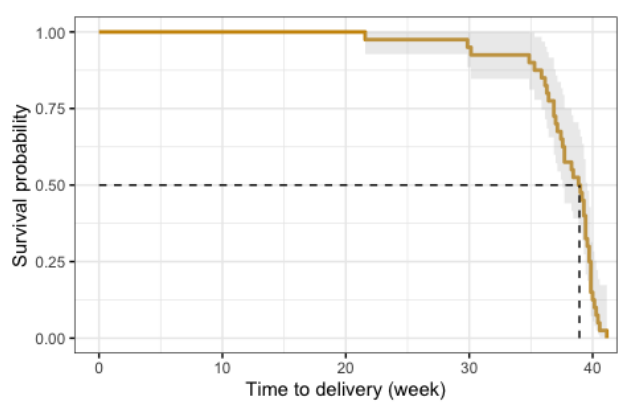}
\vspace*{-.1cm} \caption{\label{survpreg}
Kaplan-Meier survival curve for time to delivery, along with its $95\%$ confidence bands, in pregnancy data.}
\end{figure}

We utilized the joint modeling technique to analyze a public microbiome longitudinal pregnancy dataset, which was first identified in a case-control study on preterm birth outcomes conducted by  \cite{digiulio2015temporal} and  was published by  \cite{zhou2015longitudinal}.
The data we are considering includes 40 women who provided vaginal swab samples before giving birth. The study aims to explore the association between longitudinal Prevotella count outcomes (which represents the most heritable bacterial group in the vaginal microbiome and is associated with body mass index and hormonal balance) and  time to delivery.
As shown in the bar chart of the Prevotella levels  in Figure \ref{barpreg}, there is evidence of zero-inflation in these data. The score test for assessing zero-inflation in the data rejects the hypothesis of no zero-inflation, with a p-value of $<0.001$. The longitudinal model was adjusted for the log library size of each sample as a offset term. The dataset includes information on the gestational week of collection (denoted by "s"), history of preterm births, preeclampsia, race/ethnicity, and household income level, which are considered as explanatory variables. Time to delivery is considered as the event time.
Figure \ref{survpreg} displays the estimated survival curve of the time to delivery, along with its $95\%$ confidence bands, for the pregnancy data.\\
For the joint model, as demonstrated in previous studies by  \cite {baghfalaki2021approximate} and  \cite{luna2020joint}, the only significant factor influencing the mean of the response is Preeclampsia. Therefore, we consider the following linear predictors for the rate, probability, and survival models, respectively.
\begin{eqnarray*}
 \log (\eta ({s_{ij}})) = {\beta _{10}}+{\rm{offset}}(s_{ij})  + {\beta _{11}}{s_{ij}}+{\beta _{12}}Preeclampsia_i+ {b_{i1}},
 \end{eqnarray*}
\begin{eqnarray*}
{\rm{logit}}(\pi ({s_{ij}})) = \beta _{20}  + \beta _{21} {s_{ij}} + \beta _{22} Preeclampsia_i+ {b_{i2}},
 \end{eqnarray*}
  \begin{eqnarray*}
\lambda({t_i}) &=& {\lambda_0}({t_i})\exp \{ {\alpha _{0}} + {\alpha _{1}}History_{i} +  {\alpha _{2}}Preeclampsia_i \\
&+&    {\alpha _{3}} Race_i+ {\alpha _{4}}Income_i+ \gamma_1 {b_{i1}} + \gamma_2 {b_{i2}} \},
   \end{eqnarray*}
where $History$, $Preeclampsia$, $Race$, and $Income$ are used to represent the history of preterm births, preeclampsia, race (non-Hispanic whites), and high income, respectively. Also, $\rm{offset}(s)$ represents the offset term at time $s$.
For the Bayesian implementation of the model, we consider 200,000 iterations with a thinning factor of 20, and half of them are considered as pre-convergence burn-in.\\
The analysis, conducted using ZINB joint models, yields the findings presented in Table \ref{applipreg}. The Gelman-Rubin diagnostic statistic confirms the convergence of MCMC for all parameters, ensuring the reliability of the results.
Notably, the over-dispersion parameter emerges as significant. Moreover, the estimated covariance matrix of random effects reveals heterogeneity in the responses, with two random effects displaying a negative correlation.
As time progresses, there is an observable increase in the average of the responses, accompanied by a decrease in the likelihood of observing zero occurrences. This trend suggests a dynamic evolution of the studied phenomena over time.
Furthermore, the presence of preeclampsia is associated with an increase in the prevalence of Prevotella, as well as a decrease in the probability of zero occurrences, indicating its influence on the outcome.
Additionally, the history of preterm births emerges as a significant factor affecting the time to delivery, underscoring its importance in predicting pregnancy outcomes. The association parameters for the mean and the probability model are $\hat{\gamma}_1= 0.170 (-0.153,0.536)$ and $\hat{\gamma}_2= 0.099 (-0.231,0.478)$, respectively.  These findings suggest that both the occurrence and levels of Prevotella are not significantly correlated with the timing of delivery.
The complete R code for analyzing this data can be found at
\url{https://github.com/tbaghfalaki/JM-with-BUGS-and-JAGS/blob/main/Zero_inflation/code.R}.


\begin{table}
 \caption{\label{applipreg}
Results of the ZINB joint model on pregnancy data: Parameter estimates (Est.), Standard deviation (SD.), lower bound of 95\% credible
interval ($2.5\%$), upper bound of 95\% credible interval ($97.5\%$), and Gelman-Rubin statistics ($\hat{R}$).
 }
\centering
\footnotesize
 \begin{tabular}{cccccccc}
\hline 
&  &  & \multicolumn{5}{c}{Posterior statistics} \\ 
\cline{4-8}
&  & \multicolumn{1}{l}{Parameter} & Est. & SD. & $2.5\%$ & $97.5\%$ & $\widehat{R}$ \\ 
\hline
\multicolumn{4}{l}{\textit{Mean model}}   &  &   \\
  & \multicolumn{1}{l}{Intercept} & $\beta_{10}$ & -5.483 & 0.625 & -6.741 & -4.277 & 1.001  \\ 
  & \multicolumn{1}{l}{Slope}        & $\beta_{11}$ & 0.045 & 0.014 & 0.019 & 0.073 & 1.001  \\ 
  & \multicolumn{1}{l}{Preeclampsia}        & $\beta_{12}$ & 2.171 & 0.992 & 0.303 & 4.161 & 1.004  \\ 
    & \multicolumn{1}{l}{Dispersion}     &  $r$ & 0.265 & 0.033 & 0.203 & 0.333 & 1.001  \\ 

\\
\multicolumn{4}{l}{\textit{Probability model}}   &  &   \\
  & \multicolumn{1}{l}{Intercept} & $\beta_{20}$ & -1.098 & 0.658 & -2.396 & 0.154 & 1.001   \\ 
  & \multicolumn{1}{l}{Slope}        & $\beta_{21}$ & -0.034 & 0.016 & -0.065 & -0.004 & 1.001   \\ 
  & \multicolumn{1}{l}{Preeclampsia}        & $\beta_{22}$ & -1.225 & 1.121 & -3.501 & 0.934 & 1.005   \\ 
\\
\multicolumn{4}{l}{\textit{Time-to-event model}}   &  &   \\
  & \multicolumn{1}{l}{Intercept} & $\alpha_{0}$ & -80.302 & 12.717 & -103.242 & -56.152 & 1.015   \\ 
  & \multicolumn{1}{l}{History}        & $\alpha_{1}$ & 1.092 & 0.536 & 0.024 & 2.104 & 1.002  \\ 
  & \multicolumn{1}{l}{Preeclampsia}        & $\alpha_{2}$ & -0.076 & 0.495 & -1.075 & 0.892 & 1.001   \\ 
  & \multicolumn{1}{l}{Race}        & $\alpha_{3}$ & 0.072 & 0.454 & -0.822 & 0.949 & 1.001   \\ 
  & \multicolumn{1}{l}{Income}        & $\alpha_{4}$ & 0.835 & 0.488 & -0.125 & 1.781 & 1.001   \\ 
  & \multicolumn{1}{l}{Association with mean }        & $\gamma_1$ & 0.170 & 0.175 & -0.153 & 0.536 & 1.001  \\ 
  & \multicolumn{1}{l}{Association with probability}      & $\gamma_2$ & 0.099 & 0.181 & -0.231 & 0.478 & 1.001   \\ 
& \multicolumn{1}{l}{Random intercept's variance for mean model }  & $D_{11}$ & 5.258 & 1.586 & 2.900 & 9.233 & 1.001   \\ 
& \multicolumn{1}{l}{Random intercept's variance for probability model}  & $D_{22}$ & 5.306 & 2.076 & 2.551 & 10.571 & 1.002   \\ 
& \multicolumn{1}{l}{Random effect's covariance}  & $D_{12}$ & -3.589 & 1.386 & -6.979 & -1.525 & 1.001   \\ 
  & \multicolumn{1}{l}{Shape}     &  $\nu$ & 21.810 & 3.454 & 15.251 & 28.050 & 1.026   \\ 
    \hline
 \end{tabular}
\vspace*{6pt}
\end{table} 
\FloatBarrier

\section{JM of longitudinal marker and time-to-event cure outcome }

Classical survival analysis is based on the assumption that all individuals will develop the event of interest. In previous models, this assumption is reflected in the use of a so-called proper survival function which is equal to $1$ when $t=0$ and is equal to $0$ when $t$ tends towards infinity. However, the later result is not always realistic in many applications where the survival function will never tend to 0. Visually, this means that the survival function tends at a certain proportion after a certain time, reaching a plateau if the follow-up time is long enough (see Figure \ref{fig:aids_cure}). This proportion represents the fraction of individuals said to be "cured" or not susceptible to develop the event of interest. For instance, thanks to the large progress made in medicine in recent decades, certain types of cancer diagnosed in the early stages (such as melanoma, thyroid cancer or testicular cancer) have become curable, and recurrence of the disease will never be observed in some patients. They are therefore considered cured. Of course, depending on the event of interest considered, the notion of "cured" is a delicate one. Therefore, subjects who will never experience the event are more likely to be qualified as "long-term survivors" (versus short-term survivors), or also unsusceptible (versus susceptible) to developing the event. This is particularly the case when the event of interest is death caused by disease.\\
To deal with the presence of cure fraction, two main families of models have emerged in the literature  \cite{amico_2018_review}: the promotion time cure model  \cite{tsodikov_ptc_1998} and the mixture cure model  \cite{farewell_mcm_1982}. 
In the promotion time cure model, the idea is to consider an improper survival function by imposing that the latter tends towards a non-zero proportion corresponding to the fraction of cured patients. 
Conversely, the mixture cure model is a mixture model that assumes the existence of two sub-populations. The first, defined by individuals who will never experience the event and for whom no survival function is defined, and the second grouping individuals who will develop the event sooner or later and for whom a proper survival function is defined.\\
Since the 2000s, joint models have been proposed to take into account a cure fraction in the population based either on a mixture cure model  \cite{law_joint_2002,yu_joint_2004,yu_individual_2008,pan_joint_2014,barbieri_2020} or a promotion time cure model  \cite{brown_bayesian_2003,martins_2017_aids}. 
We will focus only on the ones based on mixture cure model for dealing with the cure fraction, as they correspond to a special case of joint latent class models which are not directly covered in this tutorial.

\subsection{Model specification}\label{sec:JCM}
Let $U$ denote the uncured status, where $U=1$ represents an uncured subject and $U=0$ represents a cured subject. The mixture cure model assumes that the survival function for the whole population is defined as a mixture of survival functions defined by:
\begin{equation}\label{eq:mcm}
S\left( t \right) = \left( 1- p \right) + p S_{\text{u}}\left( t \vert U=1 \right)
\end{equation}
where $S_{\text{u}}\left( t \vert U=1 \right)$ is the proper survival function associated with a proportion $p$ of uncured subjects, while the one associated with the proportion $(1-p)$ of cured subjects is implicitly fixed at 1 whatever $t\in\mathbb{R}^{+}$. As for the JM for zero-inflated longitudinal marker presented in Section \ref{ZIJM1}, the proportion $p$ of the mixture is usually not considered as a parameter as in classical mixture models but is rather modeled based on the available data. The probability $p_i$ of being uncured for the \textit{i}-th subject is classically modeled through a logistic regression model, called the "incidence model", such as:
\begin{eqnarray}\label{eq:mcm_incidence}
p_i = Pr\left(U_i=1\vert {W_{1i}}={w_{1i}}\right) & = & \frac{\exp\left({w_{1i}}^{\top}{\xi}\right)}{1+ \exp\left({w_{1i}}^{\top}{\xi}\right)},
\end{eqnarray}
with ${\xi}$ the parameter vector associated with the vector of time-independent covariates ${w_{1i}}$ and the first component of these two vectors corresponds to the intercept of the logistic regression model. 
Censoring prevents the event from being observed in all uncured individuals. It is therefore impossible to identify cured individuals from uncured ones for those where the event has not been observed. In fact, the binary variable $U_i$ is observed and equals $1$ only if subject $i$ has developed the event of interest.
In this way, a cure model can be seen as a latent class model in which the class membership random variable is partially known.\\
Extending the mixture cure model to the joint model means extending the mixture to the longitudinal sub-model as well. As with the survival model, the considered mixed model is thus defined conditionally on the cure status, leading to introduce class-specific parameters. Considering a linear mixed model, the longitudinal sub-model is:
\begin{eqnarray*}\label{eq:mcm_LMM}
Y_{i}(s_{ij}) \vert_{_{U_i=u}} & = & {x}^{\top}_{i}(s_{ij}){\beta}_u + {z}^{\top}_{i}(s_{ij}){b_i} + \varepsilon_{i}(s_{ij}) ,\quad u=0,1 
\end{eqnarray*}
where ${\beta}_u$ is the class-specific vector of fixed effects associated with the design vector ${x}_{i} (s_{ij})$, ${b_i}\sim\mathcal{N}({0},{D}_u)$ is the vector of subject-specific random effects associated with the design vector ${z}_{i} (s_{ij})$, ${D}_u$ the class-specific covariance matrix of random effects, and $\varepsilon_{i}(s_{ij})\sim\mathcal{N}(0,\sigma^2_u)$ is the measurement error term assumed to be independent of random effects. \\
Finally, the survival sub-model is only defined for uncured fraction and also depends on the shared association:  
\begin{eqnarray*}\label{eq:MCM_haz}
\lambda\left(t\vert U_i=1, \mathcal{H}_i\left(t\right) \right) & = & \lambda_0(t)\exp\left( {w_{2i}}^{\top}{\alpha} + g\left( \mathcal{H}_i\left(t\right), {b}_i,{\gamma} \right) \right),
\end{eqnarray*}
where ${w_{2i}}$ is the vector of time-independent covariates associated with the vector of parameters ${\alpha}$, and $\lambda_0(t)$ is the baseline hazard function. The two vectors of baseline covariates ${w_{1i}}$ and ${w_{2i}}$ may or may not overlap with each other. The associated survival function is then defined by:
\begin{eqnarray*}\label{eq_surv_JLSCM}
S\left( t\vert U_i=1, \mathcal{H}_i\left(t\right) \right) & = & \exp\left\{ -\int_{0}^{t}  \lambda\left(s\vert U_i=1, \mathcal{H}_i\left(s\right) \right) ds \right\}.
\end{eqnarray*}
As for the other models defined in the previous sections, the posterior distribution is given by $\pi\left({\theta},{b},{u} \vert {y},{t},{\delta}\right) \propto \mathcal{L}\left( {y},{t},{\delta} \vert {b},{u},{\theta} \right) \pi\left({b},{u},{\theta}\right)$ where $\mathcal{L}$ denotes the likelihood and ${\theta}=\left({\alpha},{\gamma}, {\xi}, \lambda_0, {\beta_0}, {\beta_1}, {D_0},{D_1},\sigma_0^2,\sigma_1^2\right)^{\top}$ is the parameter vector. Assuming that $\pi\left({b},{u},{\theta}\right)=\pi\left({b}\vert{u},{\theta}\right)\pi\left({u}\vert{\theta}\right)\pi\left({\theta}\right)$, the posterior distribution is:
\begin{eqnarray}\label{eq_likelihood_JCM}
\pi\left({\theta},{b},{u}\vert {y},{t},{\delta}\right) & \propto  & \prod^{n}_{i=1} \left\{ p_i  \lambda\left(t_i\vert u_i, \mathcal{H}_i\left(t\right) \right) S\left(t_i\vert u_i, \mathcal{H}_i\left(t\right) \right) \prod^{n_i}_{j=1} f\left( y_{ij} \vert {b_i},u_i \right) \phi\left( {b_i}\vert u_i\right) \right\}^{\delta_i u_i} \label{eq:li_du}\\
&& \times\left\{ p_i S\left(t_i\vert u_i, \mathcal{H}_i\left(t\right) \right)\prod^{n_i}_{j=1} f\left( y_{ij} \vert {b_i},u_i \right) \phi\left( {b_i}\vert u_i \right) \right\}^{(1-\delta_i)u_i} \label{eq:li_u}\\
&& \times\left\{ \left(1-p_i\right) \prod^{n_i}_{j=1} f\left( y_{ij} \vert {b_i}, u_i\right) \phi\left( {b_i}\vert u_i\right) \right\}^{(1-\delta_i)(1-u_i)} \label{eq:li_cure}\\ 
&& \times\quad\pi\left({\theta}\right)\label{eq:li_prior}
\end{eqnarray}
with $f\left(y_{ij}\vert {b_i}, u_i\right)$ is normal density function with $y_{ij}=y_{i}\left(s_{ij}\right)$, and $\phi$ denotes the zero-mean normal density function for subject-specific random effects. The line \eqref{eq:li_du} corresponds to the contribution of uncured individual who experienced the event, the line \eqref{eq:li_u} corresponds to the contribution of uncured and censored individual, the line \eqref{eq:li_cure} corresponds to the contribution of cured individual, and the line \eqref{eq:li_prior} the prior distribution of ${\theta}$.\\
Concerning prior distributions, we assume normal distribution for regression parameters and inverse-gamma distribution for positive parameters as variance parameters. For the covariance matrix, it is classical to assume an inverse Wishart distribution using BUGS. Given a Weibull baseline hazard function, the prior distributions are
\begin{eqnarray*}
{\beta_u} \sim \mathcal{N}\left( {\mu_\beta}, {\Sigma_\beta} \right),
\quad {\sigma_u} \sim \mathcal{IG}\left( {\zeta_{\sigma,1}}, {\zeta_{\sigma,2}}\right),
\quad {D_u} \sim \mathcal{IW}\left( {\Omega}, {\omega} \right), 
\quad \text{for } u=0,1 && \\
{\alpha} \sim \mathcal{N}\left( {\mu_\alpha}, {\Sigma_\alpha} \right),
\quad {\gamma} \sim \mathcal{N}\left( {\mu_\gamma}, {\Sigma_\gamma} \right), 
\quad {\xi} \sim \mathcal{N}\left( {\mu_\xi}, {\Sigma_\xi} \right),
\quad \nu \sim \mathcal{G}\left( {\zeta_{\nu,1}}, {\zeta_{\nu,2}}\right).&&
\end{eqnarray*}
where the scale parameter $\lambda_0$ of the Weibull hazard function corresponds to the intercept parameter of the latency submodel such as $\alpha_1=log\left(\lambda_0\right)$.

\subsection{Model implementation}

To present the joint cure model implementation, we consider: (1) the incidence submodel defined in equation \eqref{eq:mcm_incidence}, the survival submodel assuming a Weibull baseline hazard defined in section 2.4.2, and a simple linear subject-specific trajectory for the longitudinal submodel without assuming additional covariate. Therefore, the linear mixed model is
\begin{equation}\label{eq:linearLMM}
\begin{array}{rcl}
Y_{i}(s_{ij}) \vert_{_{U_i=u}} & = & \beta_{u0} + \beta_{u1} s_{ij} + b_{0i} + b_{1i} s_{ij} + \varepsilon_{i}(s_{ij}) ,\quad u=0,1 \\
& =& \mu_{iu}(s_{ij}) + \varepsilon_{i}(s_{ij}),
\end{array}
\end{equation}
with ${b_i}\sim\mathcal{N}\left(0, D_u \right)$ and $\varepsilon_{i}(s_{ij})\sim\mathcal{N}\left(0, \sigma^2_u \right)$ for $u=0,1$. \\
Listing \ref{lst:JCM_bugs} shows the BUGS model associated with the complete-data likelihood presented in equation \eqref{eq_likelihood_JCM} and given the longitudinal submodel in equation \eqref{eq:linearLMM}. In this model, the incidence part of the model is new (listing \ref{lst:JCM_bugs}, lines 11-15), in which it is necessary to define the class index for class-specific parameters given class membership variable $U$ (listing \ref{lst:JCM_bugs}, line 15). Note that class-specific parameters are only defined in the longitudinal sub-model, as the survival model is only defined for subjects susceptible to develop the event. Moreover, we find the survival sub-model defined in previous listings (listing \ref{lst:JCM_bugs}, lines 16-29). The only modification is that the individual contribution to the likelihood of the individual for the survival part is defined given by both the censoring indicator and the class membership $u$ (listing \ref{lst:JCM_bugs}, line 33).

\begin{lstlisting}[caption=BUGS model for the joint cure model defined in section \ref{sec:JCM}.,label={lst:JCM_bugs}]
model{
    for (i in 1:n) { 
      # Individual contribution
      # Longitudinal submodel: Distribution of Y
      for(j in id_row[i]:(id_row[i+1]-1)){
        y[j] ~ dnorm(mu[j], prec.tau2[index[i]])
        mu[j] <- inprod(beta[index[i], 1:ncX], X[j, 1:ncX]) + inprod(b[i, 1:ncZ], Z[j, 1:ncZ])
      }
      # Longitudinal submodel: Distribution of random effects
      b[i, 1:ncZ] ~ dmnorm(mu0, prec.mat[index[i], , ])
      # Incidence submodel
      logit(pi[i]) <- inprod(xi[1:ncW1], W1[i, 1:ncW1])
      u[i] ~ dbern(pi[i])
      # define the index for class specific parameters
      index[i] <- 1*equals(u[i],1) + 2*equals(u[i],0)
      # Latency submodel
      A0[i] <- gamma * ( beta[index[i], 1] + b[i, 1] )
      A1[i] <- gamma * ( beta[index[i], 2] + b[i, 2] )
      # Hazard function
      aW2[i] <- inprod(alpha[1:ncW2], W2[i, 1:ncW2])
      haz[i]<- nu * pow(st[i], nu-1) * exp(aW2[i] + A0[i] + A1[i] * st[i])
      # Cumulative hazard function
      for (k in 1:K) {
        # Scaling Gauss-Kronrod / Legendre quadrature
        xk1[i, k] <- (xk[k]+1)*st[i]/2
        wk1[i, k] <- wk[k]*st[i]/2
        # Hazard function at Gauss-Kronrod nodes
        chaz[i, k] <- nu*pow(xk1[i, k], nu-1) * exp(aW2[i] + A0[i] + A1[i]*xk1[i, k] )
      }
      # Log-survival function with Gauss-Kronrod / Legendrer requadrature
      logSurv[i] <- -inprod( wk1[i, 1:K], chaz[i, 1:K] )
      # Definition of the survival log-likelihood using zeros trick
      logL[i] <- u[i]*log(pi[i]) + u[i]*delta[i]*log(haz[i]) + u[i]*logSurv[i] + (1-delta[i])*(1-u[i])*log(1-pi[i])
      phi[i] <- 100000 - logL[i]
      zeros[i] ~ dpois(phi[i])
    }
    # contribution to prior
    # Longitudinal part
    prec.tau2[1] ~ dgamma(0.01, 0.01)
    prec.tau2[2] ~ dgamma(0.01, 0.01)
    beta[1,1:ncX] ~ dmnorm(priorMean.beta[], priorTau.beta[, ])         
    beta[2,1:ncX] ~ dmnorm(priorMean.beta[], priorTau.beta[, ])          
    prec.mat[1, 1:ncZ, 1:ncZ] ~ dwish(priorR.prec.mat[, ], priorK.prec.mat)  
    prec.mat[2, 1:ncZ, 1:ncZ] ~ dwish(priorR.prec.mat[, ], priorK.prec.mat)  
    # Incidence priors
    xi[1:ncW1] ~ dmnorm(priorMean.xi[], priorTau.xi[, ])
    # Latency/survival priors
    nu ~ dgamma(0.01, 0.01)
    alpha[1:ncW2] ~ dmnorm(priorMean.alpha[], priorTau.alpha[, ])
    gamma ~ dnorm(0, precision)
    # determinetic nodes
    sigma2[1] <- 1/prec.tau2[1]
    sigma2[2] <- 1/prec.tau2[2]
    covariance.b[1, 1:ncZ, 1:ncZ] <- inverse(prec.mat[1, 1:ncZ, 1:ncZ])
    covariance.b[2, 1:ncZ, 1:ncZ] <- inverse(prec.mat[2, 1:ncZ, 1:ncZ])
  }
\end{lstlisting}

\subsection{Application to AIDS data}

\begin{figure}
\centering
\includegraphics[width=10cm]{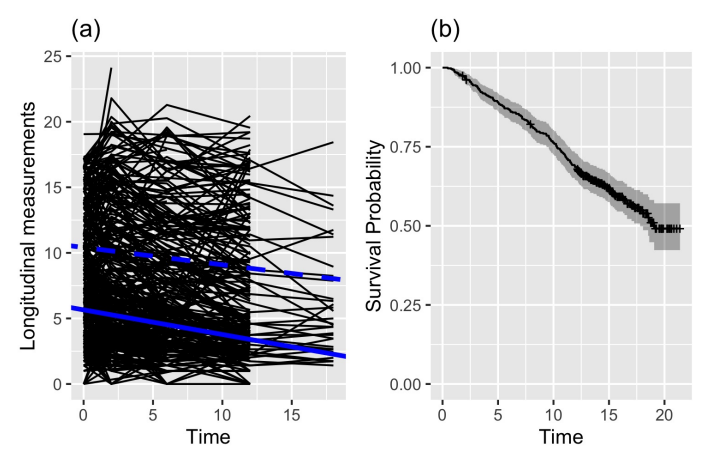}
\caption{Spaghetti plot for CD4 measurements on (a) with mean trajecotries for long-term and short-term survivors representing by dotted and solid lines, respectively ; and Kaplan-Meier surival estimation plot on (b).}\label{fig:aids_cure}
\end{figure}

To illustrate an application for joint modeling of longitudinal measurements and time-to-event cure outcomes, we considered AIDS data collected during a randomized clinical trial  \cite{abrams_1994} designed to compare the efficacy and safety of two antiretroviral drugs in treating HIV-infected patients. The 467 patients were included after a failure or an intolerance of previous zidovudine (AZT) therapy.
Freely available in the \texttt{JMbayes} package, these data include both survival data and longitudinal data. The longitudinal outcome is the square root of the CD4 lymphocyte count, measured at baseline, 2, 6, 12 and 18 months (see spaghetti plot (a) of Figure \ref{fig:aids_cure}). For survival data, the event of interest is the death and the associated Kaplan-Meier curve (Figure \ref{fig:aids_cure}.(b)) shows the characteristic plateau at the end of follow-up indicating the presence of a fraction of long-term survivors.
In addition, four binary explanatory variables are also available: the treatment (ddI vs ddC), the gender (male vs female), the previous opportunistic infection (AIDS vs no AIDS at study entry) and the AZT indicator (failure vs intolerance).

\FloatBarrier

In the application, we considered the longitudinal submodel defined in equation \eqref{eq:linearLMM}, and all the four explanatory variables in both the incidence and latency submodels:
\begin{equation*}
\begin{array}{lcl}
\text{logit} \left( p_i \right) & = &  \xi_1 + \xi_2 \text{Drug}_{i} + \xi_3 \text{Gender}_{i} + \xi_4 \text{PrevOI}_{i} + \xi_5 \text{AZT}_{i} \\
\lambda\left( t_i \right) & = &  \nu t_i^{\nu-1} \exp\left( \alpha_1 + \alpha_2 \text{Drug}_{i} + \alpha_3 \text{Gender}_{i} + \alpha_4 \text{PrevOI}_{i} + \alpha_5 \text{AZT}_{i} + \gamma \mu_i\left(t_i\right) \right)\\
\end{array}
\end{equation*}
where $t_i$ is the observed time for the $i$th patient.

To use the JAGS sampler, we need to provide both the model in BUGS syntax (given in listing \ref{lst:JCM_bugs}) and the data in a \texttt{list()} object. The data list must contain all observations (realization of response variables, matrices of explanatory variables), the set of values to define the model (object dimensions, number of individuals, individual row pointers, nodes and weights of the Gauss-Kronrod method, the vector of zeros for zeros tricks) and finally the hyper-parameters of the prior distributions. 
First, Listing \ref{lst:aids_long} presents the loading of data from \texttt{JMbayes} package and the creation of the list object \texttt{jags.data} including the longitudinal contribution of data. The object \texttt{id\_row} (lines 13-16, Listing \ref{lst:aids_long}) is a vector in which the element \texttt{id\_row[i]} returns the row of the first observation for the $i$th patient. Note that the vector \texttt{y} (line 8, Listing \ref{lst:aids_long}) has to be sorted by $id$.

\begin{lstlisting}[ caption={R code to load AIDS data and creation of a list object with the data associated with the longitudinal part.}, label={lst:aids_long}]
# load data
data("aids", package = "JMbayes2")
data.long <- aids
data("aids.id", package = "JMbayes2")
data.id <- aids.id
# creation of jags data list starting with longitudinal part
jags.data <- list(
  y = data.long$CD4,
  X = cbind(1, data.long$obstime),
  Z = cbind(1, data.long$obstime),
  ncX = 2,
  ncZ = 2,
  id_row = as.vector(c(1, 
                       1+cumsum(tapply(as.integer(data.long$patient),
                                       as.integer(data.long$patient),
                                       length)))))
\end{lstlisting}
Then, Listing \ref{lst:aids_surv} presents the addition of data associated with the survival part to the list \texttt{jags.data}. This part includes number $n$ of patients (line 14), the vector of observed times (line 15), the vector of death indicator (line 16), the design matrix of the linear predictor with its column number (lines 17-22), information about the 15 nodes and 15 weights for approximating the integral in survival function using Gauss-Kronrod method (lines 2-6 and lines 23-25), and the zero vector for the zeros-trick (lines 10,11,26). 
\begin{lstlisting}[ caption={R code to update the \texttt{jags.data} by including data associated with the survival part.}, label={lst:aids_surv}]
# update jags data with zeros trick
zeros <- numeric(length(data.id$Time))
# update jags data with the survival data
jags.data <- c(jags.data, 
               list(n = length(data.id$Time),
                    st = data.id$Time,
                    delta = data.id$death,
                    W2 = cbind(1, 
                               data.id$drug, 
                               data.id$gender, 
                               data.id$prevOI, 
                               data.id$AZT),
                    ncW2 = 5,
                    wk = wk, # Gauss-Kronrod weights
                    xk = sk # Gauss-Kronrod nodes
                    K = length(wk), 
                    ),
                    zeros = zeros)
\end{lstlisting}
In addition, Listing \ref{lst:aids_inc} presents the addition to the list of data associated with the incidence part. 
Concerning class membership information, we need to create a partially observed vector \textit{u} of one corresponding to the observed censoring indicator vector $\delta$ in which the elements equal to zero (associated with censoring patients) are filled in as \texttt{NA} (Listing \ref{lst:aids_inc}, lines 1-3 and 15). To improve the model estimation, the zero-tail constraint is classically used in cured models by assuming as cured all subjects with a censoring time greater than the largest observed event time (Listing \ref{lst:aids_inc}, lines 5-6). 
The zero-tail constraint should only be considered when the tracking time is sufficiently long and when the censoring time is not an administrative censoring time for the majority of censored individuals. In the application, 188 deaths were observed and only 26 patients out of 467 were censored after the longest event time observed.
As for the survival part, we provided the design matrix of the linear predictor with its column number (lines 9-14).
\begin{lstlisting}[ caption={R code to update the \texttt{jags.data} by including data associated with the incidence part.}, label={lst:aids_inc}]
# Define the partially latent variable U for class membership
u <- data.id$death
u[which(u==0)] <- NA
# zero-tail constraint
u[which(data.id$death==0 & 
          data.id$Time>max(data.id$Time[which(data.id$death==1)]))] <- 0
# update jags data with the incidence data
jags.data <- c(jags.data, 
               list(W1 = cbind(1, 
                               data.id$drug, 
                               data.id$gender, 
                               data.id$prevOI, 
                               data.id$AZT),
                    ncW1 = 5,
                    u = u))
\end{lstlisting}
Finally, Listing \ref{lst:aids_prior} presents the addition to the list of fixed parameters of prior distributions based. In this listing, we assume vague priors as previously discussed by considering a common precision fixed to $0.01$ to define the vague priors (line 2). 
\begin{lstlisting}[ caption={R code to update the \texttt{jags.data} by including information about the prior definition.}, label={lst:aids_prior}]
# Choice of precision
precision <- 0.01
# prior parameters for longitudinal submodel
priorMean.beta <- as.numeric(rep(0, jags.data$ncX))
priorTau.beta <- diag(rep(precision, length(priorMean.beta)))
mu0 <- rep(0, jags.data$ncZ)
priorR.prec.mat <- diag(rep(precision, jags.data$ncZ))
priorK.prec.mat <- jags.data$ncZ
# prior parameters for incidence submodel
priorMean.xi <- as.numeric(rep(0, jags.data$ncW1))
priorTau.xi <- diag(rep(precision, length(priorMean.xi)))
# prior parameters for latency submodel
priorMean.alpha <- as.numeric(rep(0, jags.data$ncW2))
priorTau.alpha <- diag(rep(precision, length(priorMean.alpha)))
# update jags data with hyper parameters
jags.data <- c(jags.data, 
               list(precision = precision, 
                    priorMean.beta = priorMean.beta,
                    priorTau.beta = priorTau.beta,
                    priorMean.xi = priorMean.xi,
                    priorTau.xi = priorTau.xi,
                    priorMean.alpha = priorMean.alpha,
                    priorTau.alpha = priorTau.alpha,
                    mu0 = mu0,
                    priorR.prec.mat = priorR.prec.mat,
                    priorK.prec.mat = priorK.prec.mat))
\end{lstlisting}
Once the data list is complete, you need to specify the parameters and/or random variables whose posterior samples you want to save (Listing \ref{lst:aids_jagsui}, lines 1-2). And finally, to call JAGS sampler using \texttt{jags} function from \texttt{jagsUI} package, note that \texttt{jags.model} is a string defined by the model in Listing \ref{lst:JCM_bugs}, given between quotation marks. The choice of number of chains, number of iterations, burn-in phase, thinning rate is the one used to apply the joint cure model on the application.
\begin{lstlisting}[ caption={R code to specify the parameters and/or random variables to save, and to call JAGS sampler using \texttt{jags} function from \texttt{jagsUI} package.}, label={lst:aids_jagsui}]
# parameters to save in the sampling step
parms_to_save <- c("beta", "sigma2", "xi", "alpha", "gamma", "nu", "covariance.b", "b")
# using jagsUI
out_jags = jagsUI::jags(data = jags.data,
                        parameters.to.save = parms_to_save,
                        model.file = textConnection(jags.model),
                        n.chains = 3,
                        parallel = FALSE,
                        n.adapt = NULL,
                        n.iter = 200000,
                        n.burnin = 50000,
                        n.thin = 1,
                        DIC = F)
\end{lstlisting}
Table \ref{tab:estimation_aids} presents some statistics from posterior samples. First, the Gelman-Rubin statistics indicate the convergence of MCMC for all parameters. 
For the incidence sub-model, only having the previous opportunistic infection impacts the probability of short-term survival. In fact, patients previously infected with AIDS are more likely to be short-term survivors ($\hat{\xi}_3=4.354\; (3.667,5.141)$). 
For the latency sub-model, the risk of death is lower for patients previously infected with AIDS ($\hat{\alpha}_3=-1.320\;(-1.859,-0.756)$). As expected, the risk of death decreases with increasing CD4 count ($\hat{\gamma}=-0.263\;(-0.353,-0.181)$).
For the longitudinal sub-model, there is a difference of mean trajectory for CD4 counts between long-term and short-term survivors (see Figure \ref{fig:aids_cure}). Long-term survivors have on average twice as many CD4s as short-term survivors ($\hat{\beta}_{0,0}=10.413\;(9.566,11.228)$ and $\hat{\beta}_{1,0}=5.656\;(5.207,6.107)$, respectively). For both groups, CD4 counts tends to decrease over time ($\hat{\beta}_{0,1}=-0.132\;(-0.177,-0.088)$ and $\hat{\beta}_{1,1}=-0.188\;(-0.227,-0.150)$, respectively). Furthermore, there was greater variability among long-term survivors, both in terms of random intercept variability and error variability ($\hat{D}_{0,11}=22.268\;(17.101,28.733)$ and $\hat{\sigma}^2_{0}=4.974\;(4.154,5.882)$, respectively).
\begin{table}[t!]
\centering
\caption{Posterior statistics including posterior mean (Est.), posterior standard deviation (SD.), bound of the $95\%$ credible interval (posterior quantiles with orders $2.5\%$ and $97.5\%$) and the Gelman-Rubin statistics ($\widehat{R}$). Results are obtained using 150000 draws from the MCMC procedure after a burning of 50000 iterations.}\label{tab:estimation_aids}
\scalebox{0.8}{
\begin{tabular}{cccccccc}
\hline 
&  &  & \multicolumn{5}{c}{Posterior statistics} \\ 
\cline{4-8}
&  & \multicolumn{1}{l}{Parameter} & Est. & SD. & $2.5\%$ & $97.5\%$ & $\widehat{R}$ \\ 
\hline
\multicolumn{4}{l}{\textit{Incidence model}}   &  &   \\
  & \multicolumn{1}{l}{Intercept} & $\xi_1$ & -6.266 & 0.960  & -8.167  & -4.407 & 1.001  \\ 
  & \multicolumn{1}{l}{Drug$_{\text{ddI}}$} & $\xi_2$ & 0.384  &  0.248  &  -0.095  &  0.874 & 1.003 \\ 
  & \multicolumn{1}{l}{Gender$_{\text{male}}$}  & $\xi_3$ & -0.097  & 0.409  & -0.909  & 0.692 & 1.001  \\ 
  & \multicolumn{1}{l}{PrevOI$_{\text{AIDS}}$}  & $\xi_4$ & 4.354  &  0.374  &  3.667  &  5.141 &  1.004  \\ 
  & \multicolumn{1}{l}{AZT$_{\text{failure}}$}  & $\xi_5$ & -0.004  & 0.408  & -0.819  & 0.787 & 1.003  \\ 
\\
\multicolumn{2}{l}{\textit{Latency model}} &   &   &  &   \\  
  & \multicolumn{1}{l}{Shape}     &  $\nu$ & 1.464 & 0.103 & 1.267 & 1.673 & 1.001  \\ 
  & \multicolumn{1}{l}{Intercept} & $\alpha_1=log(\lambda_0)$ & -0.852 & 0.785 & -2.409 & 0.656 & 1.001  \\ 
  & \multicolumn{1}{l}{Drug$_{\text{ddI}}$}  & $\alpha_2$ & 0.178 & 0.156 & -0.126 & 0.488 & 1.000  \\ 
  & \multicolumn{1}{l}{Gender$_{\text{male}}$}  & $\alpha_3$ & -0.142 & 0.269 & -0.653 & 0.406 & 1.001  \\ 
  & \multicolumn{1}{l}{PrevOI$_{\text{AIDS}}$}  & $\alpha_4$ & -1.320 & 0.280 & -1.859 & -0.756 & 1.002  \\ 
  & \multicolumn{1}{l}{AZT$_{\text{failure}}$}  & $\alpha_5$ &  0.147 & 0.169 & -0.185 & 0.480 & 1.000  \\ 
  & \multicolumn{1}{l}{Current value}           & $\gamma$ & -0.263 & 0.044 & -0.353 & -0.181 & 1.001 \\  
\\ 
\multicolumn{7}{l}{\textit{Longitudinal model for susceptible patients ($U=1$)}} \\ 
  & \multicolumn{1}{l}{Intercept}    & $\beta_{1,0}$ & 5.656 & 0.23 & 5.207 & 6.107 & 1.002 \\ 
  & \multicolumn{1}{l}{Slope}        & $\beta_{1,1}$ & -0.188 & 0.02 & -0.227 & -0.150 & 1.001 \\ 
  & \multicolumn{1}{l}{Random intercept's variance} & $D_{1,11}$ & 13.634 & 1.342 & 11.165 & 16.433 & 1.000\\ 
  & \multicolumn{1}{l}{Random slope's variance}   & $D_{1,22}$ & 0.034 & 0.009 & 0.020 & 0.054 & 1.020 \\ 
  &\multicolumn{1}{l}{Random effect's covariance} &  $D_{1,12}$ & -0.405 & 0.079 & -0.570 & -0.258 & 1.001\\ 
  & \multicolumn{1}{l}{Residual variance} & $\sigma^2_{1}$ & 2.032 & 0.155 & 1.750 & 2.356 & 1.001\\
\\ 
\multicolumn{7}{l}{\textit{Longitudinal model for unsusceptible patients ($U=0$)}} \\ 
  & \multicolumn{1}{l}{Intercept}      & $\beta_{0,0}$ & 10.413 & 0.422 & 9.566 & 11.228 & 1.001 \\ 
  & \multicolumn{1}{l}{Slope}          & $\beta_{0,0}$ & -0.132 & 0.023 & -0.177 & -0.088 & 1.001 \\ 
  & \multicolumn{1}{l}{Random intercept's variance}  & {$D_{0,11}$}     & 22.268 & 2.977 & 17.101 & 28.733 & 1.000 \\ 
  & \multicolumn{1}{l}{Random slope's variance}   & {$D_{0,22}$}     & 0.010 & 0.007 & 0.002 & 0.028 & 1.002 \\ 
  &\multicolumn{1}{l}{Random effect's covariance}  &{$D_{0,12}$}      & 0.162 & 0.109 & -0.055 & 0.377 & 1.003 \\ 
   & \multicolumn{1}{l}{Residual variance}  & {$\sigma^2_{0}$} & 4.974 & 0.441 & 4.154 & 5.882 & 1.003 \\
\hline 
\end{tabular} 
}
\end{table}
\FloatBarrier
\section{Conclusions}
The application of survival analysis in medicine and biological sciences has witnessed significant advancements, particularly through the incorporation of the Bayesian paradigm and the BUGS syntax. Alvares et al.  \cite{alvares2021bayesian} have demonstrated the versatility and applicability of this approach across various well-established models, such as Accelerated Failure Time (AFT), Proportional Hazards (PH), mixture cure, competing risks, multi-state, frailty, and joint models.\\
In the realm of clinical research and healthcare, joint modeling has emerged as an indispensable tool, facilitating the simultaneous analysis of longitudinal and survival data. This integrated approach enables researchers to explore the correlation between repeated measurements and their impact on time-to-event outcomes, fostering a deeper understanding of intricate variable relationships. Joint modeling finds diverse applications in cancer studies, cardiovascular research, HIV/AIDS studies, psychiatric investigations, neurological disorders, and recurrent events analysis.\\
Within joint modeling, the Bayesian framework stands out as a robust statistical approach, providing flexibility, coherence, and effective modeling of complex relationships. Bayesian methodologies leverage samples from the posterior distribution, enabling flexible model specification, handling of uncertainty, incorporation of prior information, and robust estimation of parameter variances.\\
This paper introduces the implementation of a Bayesian approach using BUGS, complemented by the versatile JAGS (Just Another Gibbs Sampler). JAGS offers flexibility, ease of use, an open-source nature, seamless integration with R through the \texttt{rjags} package, and a vibrant user community.\\
The paper explores advanced joint modeling techniques, leveraging the capabilities of BUGS and JAGS. This includes simulating real data, extending analyses to multiple longitudinal markers, addressing competing risks and cure time-to-event outcome, and handling zero-inflated longitudinal measurements. Additionally, proportional hazard models with varying forms of baseline hazards and diverse methods for addressing association structures between sub-models are explored.\\
Furthermore, the paper includes programs for computing posterior summaries and generating plots to assist in interpreting the results. The shared code in BUGS facilitates its application in medicine and related fields, promoting innovation and enabling new discoveries.\\
In summary, this paper presents the implementation of each model using BUGS syntax, which is executable with JAGS in the R programming language. All data, simulated and real, along with the code for various models, are accessible at \url{https://github.com/tbaghfalaki/JM-with-BUGS-and-JAGS}. Through this effort, we aim to empower researchers to navigate the complexities of joint modeling and drive advancements in survival analysis research.\\
To conclude, it is noteworthy that in BUGS, users can define prior distributions for regression coefficients, often using distributions such as the normal distribution. For instance, a low informative prior for the regression coefficient can be represented by a normal distribution with a mean of 0 and a large standard deviation. The software also supports the inverse-Wishart distribution for covariance matrices and allows users to assign a prior distribution for the residual variance term, usually an inverse-gamma distribution. Custom priors for covariance matrices are also available. In Stan, the language behind RStan, popular choices include the Lewandowski-Kurowicka-Joe (LKJ) distribution  \cite{lewandowski2009generating} for correlation matrices and the Wishart distribution for the covariance matrix. Sensitivity analysis concerning priors is essential in Bayesian regression modeling. The selection of these methods depends on user preferences, modeling requirements, and the level of control required for specifying priors. Analysts can utilize the distinctive characteristics of each tool to tailor their Bayesian modeling based on their data and prior knowledge, guaranteeing robust and reliable inference through systematic sensitivity analysis.

{\footnotesize{
\bibliographystyle{plain}
\bibliography{biblio.bib}}}

\begin{thebibliography}{10}

\bibitem{abrahamowicz1996time}
Michal Abrahamowicz, Todd MacKenzie, and John~M Esdaile.
\newblock Time-dependent hazard ratio: modeling and hypothesis testing with
  application in lupus nephritis.
\newblock {\em Journal of the American Statistical Association},
  91(436):1432--1439, 1996.

\bibitem{abrams_1994}
Donald~I. Abrams, Anne~I. Goldman, Cynthia Launer, Joyce~A. Korvick, James~D.
  Neaton, Lawrence~R. Crane, Michael Grodesky, Steven Wakefield, Katherine
  Muth, Sandra Kornegay, David~L. Cohn, Allen Harris, Roberta Luskin-Hawk,
  Norman Markowitz, James~H. Sampson, Melanie Thompson, Lawrence Deyton, and
  the Terry Beirn Community Programs for Clinical Research~on AIDS.
\newblock A comparative trial of didanosine or zalcitabine after treatment with
  zidovudine in patients with human immunodeficiency virus infection.
\newblock {\em New England Journal of Medicine}, 330(10):657---662, 1994.

\bibitem{alvares2021bayesian}
Danilo Alvares, Elena L{\'a}zaro, Virgilio G{\'o}mez-Rubio, and Carmen Armero.
\newblock Bayesian survival analysis with bugs.
\newblock {\em Statistics in Medicine}, 40(12):2975--3020, 2021.

\bibitem{amico_2018_review}
Maïlis Amico and Ingrid~Van Keilegom.
\newblock Cure models in survival analysis.
\newblock {\em Annual Review of Statistics and Its Application}, 5(1):null,
  2018.

\bibitem{austin2012generating}
Peter~C Austin.
\newblock Generating survival times to simulate cox proportional hazards models
  with time-varying covariates.
\newblock {\em Statistics in medicine}, 31(29):3946--3958, 2012.

\bibitem{baghfalaki2021approximate}
T~Baghfalaki and M~Ganjali.
\newblock Approximate bayesian inference for joint linear and partially linear
  modeling of longitudinal zero-inflated count and time to event data.
\newblock {\em Statistical Methods in Medical Research}, 30(6):1484--1501,
  2021.

\bibitem{bakka2018spatial}
Haakon Bakka, H{\aa}vard Rue, Geir-Arne Fuglstad, Andrea Riebler, David Bolin,
  Janine Illian, Elias Krainski, Daniel Simpson, and Finn Lindgren.
\newblock Spatial modeling with r-inla: A review.
\newblock {\em Wiley Interdisciplinary Reviews: Computational Statistics},
  10(6):e1443, 2018.

\bibitem{bao2022package}
Le~Bao, Jeff Gill, Tsunghan Tsai, Jonathan Rapkin, and Maintainer Le~Bao.
\newblock Package ‘superdiag’.
\newblock {\em R CRAN}, 2022.

\bibitem{barbieri_2020}
Antoine Barbieri and Catherine Legrand.
\newblock Joint longitudinal and time-to-event cure models for the assessment
  of being cured.
\newblock {\em Statistical Methods in Medical Research}, 29(4):1256---1270,
  2020.

\bibitem{bernhardt2015fast}
Paul~W Bernhardt, Daowen Zhang, and Huixia~Judy Wang.
\newblock A fast em algorithm for fitting joint models of a binary response and
  multiple longitudinal covariates subject to detection limits.
\newblock {\em Computational statistics \& data analysis}, 85:37--53, 2015.

\bibitem{bivand2015spatial}
Roger Bivand, Virgilio G{\'o}mez-Rubio, and H{\aa}vard Rue.
\newblock Spatial data analysis with r-inla with some extensions.
\newblock {\em American Statistical Association}, 2015.

\bibitem{bogaert2014iteration}
Ignace Bogaert.
\newblock Iteration-free computation of gauss-legendre quadrature nodes and
  weights.
\newblock {\em SIAM Journal on Scientific Computing}, 36(3):A1008--A1026, 2014.

\bibitem{bolstad2009understanding}
William~M Bolstad.
\newblock {\em Understanding computational Bayesian statistics}, volume 644.
\newblock John Wiley \& Sons, 2009.

\bibitem{brent2013algorithms}
Richard~P Brent.
\newblock {\em Algorithms for minimization without derivatives}.
\newblock Courier Corporation, 2013.

\bibitem{brezger2005bayesx}
Andreas Brezger, Thomas Kneib, and Stefan Lang.
\newblock Bayesx: analyzing bayesian structural additive regression models.
\newblock {\em Journal of statistical software}, 14:1--22, 2005.

\bibitem{brilleman2018}
Samuel~L Brilleman.
\newblock simjm: Simulate joint longitudinal and survival data.
\newblock {\em R package version 0.0.1, URL
  https://github.com/sambrilleman/simjm.}, 2018.

\bibitem{brilleman2021simulating}
Samuel~L Brilleman, Rory Wolfe, Margarita Moreno-Betancur, and Michael~J
  Crowther.
\newblock Simulating survival data using the simsurv r package.
\newblock {\em Journal of Statistical Software}, 97:1--27, 2021.

\bibitem{brown_bayesian_2003}
Elizabeth~R. Brown and Joseph~G. Ibrahim.
\newblock Bayesian {Approaches} to {Joint} {Cure}-{Rate} and {Longitudinal}
  {Models} with {Applications} to {Cancer} {Vaccine} {Trials}.
\newblock {\em Biometrics}, 59(3):686--693, September 2003.

\bibitem{christensen2010bayesian}
Ronald Christensen, Wesley Johnson, Adam Branscum, and Timothy~E Hanson.
\newblock {\em Bayesian ideas and data analysis: an introduction for scientists
  and statisticians}.
\newblock CRC press, 2010.

\bibitem{crowther2012simulating}
Michael~J Crowther and Paul~C Lambert.
\newblock Simulating complex survival data.
\newblock {\em The Stata Journal}, 12(4):674--687, 2012.

\bibitem{curtis2018package}
S~McKay Curtis, Ilya Goldin, and Evangelos Evangelou.
\newblock Package ‘mcmcplots’.
\newblock {\em R CRAN}, 2018.

\bibitem{diggle2002analysis}
Peter Diggle.
\newblock {\em Analysis of longitudinal data}.
\newblock Oxford university press, 2002.

\bibitem{digiulio2015temporal}
Daniel~B DiGiulio, Benjamin~J Callahan, Paul~J McMurdie, Elizabeth~K Costello,
  Deirdre~J Lyell, Anna Robaczewska, Christine~L Sun, Daniela~SA Goltsman,
  Ronald~J Wong, Gary Shaw, et~al.
\newblock Temporal and spatial variation of the human microbiota during
  pregnancy.
\newblock {\em Proceedings of the National Academy of Sciences},
  112(35):11060--11065, 2015.

\bibitem{ding2008modeling}
Jimin Ding and Jane-Ling Wang.
\newblock Modeling longitudinal data with nonparametric multiplicative random
  effects jointly with survival data.
\newblock {\em Biometrics}, 64(2):546--556, 2008.

\bibitem{elashoff2016joint}
Robert Elashoff, Ning Li, et~al.
\newblock {\em Joint modeling of longitudinal and time-to-event data}.
\newblock CRC press, 2016.

\bibitem{farewell_mcm_1982}
V.~T. Farewell.
\newblock The {Use} of {Mixture} {Models} for the {Analysis} of {Survival}
  {Data} with {Long}-{Term} {Survivors}.
\newblock {\em Biometrics}, 38(4):1041--1046, 1982.

\bibitem{fitzmaurice2008longitudinal}
Garrett Fitzmaurice, Marie Davidian, Geert Verbeke, and Geert Molenberghs.
\newblock {\em Longitudinal data analysis}.
\newblock CRC press, 2008.

\bibitem{gelman1992inference}
Andrew Gelman, Donald~B Rubin, et~al.
\newblock Inference from iterative simulation using multiple sequences.
\newblock {\em Statistical science}, 7(4):457--472, 1992.

\bibitem{geweke1992evaluating}
John Geweke.
\newblock Evaluating the accuracy of sampling-based approaches to the
  calculations of posterior moments.
\newblock {\em Bayesian statistics}, 4:641--649, 1992.

\bibitem{guo2020package}
Jiqiang Guo, Jonah Gabry, Ben Goodrich, and S~Weber.
\newblock Package ‘rstan’.
\newblock {\em URL https://cran. r―project. org/web/packages/rstan}, 2020.

\bibitem{guo2004separate}
Xu~Guo and Bradley~P Carlin.
\newblock Separate and joint modeling of longitudinal and event time data using
  standard computer packages.
\newblock {\em The american statistician}, 58(1):16--24, 2004.

\bibitem{heidelberger1983simulation}
Philip Heidelberger and Peter~D Welch.
\newblock Simulation run length control in the presence of an initial
  transient.
\newblock {\em Operations Research}, 31(6):1109--1144, 1983.

\bibitem{henderson2000joint}
Robin Henderson, Peter Diggle, and Angela Dobson.
\newblock Joint modelling of longitudinal measurements and event time data.
\newblock {\em Biostatistics}, 1(4):465--480, 2000.

\bibitem{hobert1996effect}
James~P Hobert and George Casella.
\newblock The effect of improper priors on gibbs sampling in hierarchical
  linear mixed models.
\newblock {\em Journal of the American Statistical Association},
  91(436):1461--1473, 1996.

\bibitem{ibrahim2001bayesian}
Joseph~G Ibrahim, Ming-Hui Chen, and Debajyoti Sinha.
\newblock {\em Bayesian survival analysis}, volume~2.
\newblock Springer, 2001.

\bibitem{klein2016handbook}
John~P Klein, Hans~C Van~Houwelingen, Joseph~G Ibrahim, and Thomas~H Scheike.
\newblock {\em Handbook of survival analysis}.
\newblock CRC Press, 2016.

\bibitem{kleinbaum1996survival}
David~G Kleinbaum and Mitchel Klein.
\newblock {\em Survival analysis a self-learning text}.
\newblock Springer, 1996.

\bibitem{lang2004bayesian}
Stefan Lang and Andreas Brezger.
\newblock Bayesian p-splines.
\newblock {\em Journal of computational and graphical statistics},
  13(1):183--212, 2004.

\bibitem{law_joint_2002}
Ngayee~J. Law, Jeremy M.~G. Taylor, and Howard Sandler.
\newblock The joint modeling of a longitudinal disease progression marker and
  the failure time process in the presence of cure.
\newblock {\em Biostatistics (Oxford, England)}, 3(4):547--563, December 2002.

\bibitem{leonard1978density}
Tom Leonard.
\newblock Density estimation, stochastic processes and prior information.
\newblock {\em Journal of the Royal Statistical Society: Series B
  (Methodological)}, 40(2):113--132, 1978.

\bibitem{lewandowski2009generating}
Daniel Lewandowski, Dorota Kurowicka, and Harry Joe.
\newblock Generating random correlation matrices based on vines and extended
  onion method.
\newblock {\em Journal of multivariate analysis}, 100(9):1989---2001, 2009.

\bibitem{lin2002modeling}
Haiqun Lin and Daniel Zelterman.
\newblock Modeling survival data: extending the cox model, 2002.

\bibitem{luna2020joint}
Pamela~N Luna, Jonathan~M Mansbach, and Chad~A Shaw.
\newblock A joint modeling approach for longitudinal microbiome data improves
  ability to detect microbiome associations with disease.
\newblock {\em PLoS computational biology}, 16(12):e1008473, 2020.

\bibitem{martins_2017_aids}
Rui Martins, Giovani~L. Silva, and Valeska Andreozzi.
\newblock Joint analysis of longitudinal and survival aids data with a spatial
  fraction of long-term survivors: A bayesian approach.
\newblock {\em Biometrical journal}, 59(6):1166--1183, 2017.

\bibitem{molenberghs2006longitudinal}
Geert Molenberghs and Geert Verbeke.
\newblock Longitudinal data analysis.
\newblock {\em Wiley StatsRef: Statistics Reference Online}, pages 1--28, 2006.

\bibitem{murray2023package}
James Murray and Maintainer~James Murray.
\newblock Package ‘gmvjoint’.
\newblock {\em R CRAN}, 2023.

\bibitem{ntzoufras2011bayesian}
Ioannis Ntzoufras.
\newblock {\em Bayesian modeling using WinBUGS}.
\newblock John Wiley \& Sons, 2011.

\bibitem{pan_joint_2014}
Jianxin Pan, Yanchun Bao, Hongsheng Dai, and Hong-Bin Fang.
\newblock Joint longitudinal and survival-cure models in tumour xenograft
  experiments.
\newblock {\em Statistics in Medicine}, 33(18):3229--3240, August 2014.

\bibitem{philipson2012joiner}
Pete Philipson, Peter Diggle, Ines Sousa, Ruwanthi Kolamunnage-Dona, Paula
  Williamson, and Robin Henderson.
\newblock joiner: Joint modelling of repeated measurements and time-to-event
  data.
\newblock {\em Comprehensive R Archive Network}, 2012.

\bibitem{plummer2012jags}
Martyn Plummer.
\newblock Jags version 3.3. 0 user manual, 2012.

\bibitem{plummer2006coda}
Martyn Plummer, Nicky Best, Kate Cowles, Karen Vines, et~al.
\newblock Coda: convergence diagnosis and output analysis for mcmc.
\newblock {\em R news}, 6(1):7--11, 2006.

\bibitem{raftery1992many}
Adrian~E Raftery, Steven Lewis, et~al.
\newblock How many iterations in the gibbs sampler.
\newblock {\em Bayesian statistics}, 4(2):763--773, 1992.

\bibitem{JM}
Dimitris Rizopoulos.
\newblock {JM}: An {R} package for the joint modelling of longitudinal and
  time-to-event data.
\newblock {\em Journal of Statistical Software}, 35(9):1--33, 2010.

\bibitem{rizopoulos2012joint}
Dimitris Rizopoulos.
\newblock {\em Joint models for longitudinal and time-to-event data: With
  applications in R}.
\newblock CRC press, 2012.

\bibitem{JSSv072i07}
Dimitris Rizopoulos.
\newblock The r package jmbayes for fitting joint models for longitudinal and
  time-to-event data using mcmc.
\newblock {\em Journal of Statistical Software}, 72(7):1–46, 2016.

\bibitem{rizopoulos2011bayesian}
Dimitris Rizopoulos and Pulak Ghosh.
\newblock A bayesian semiparametric multivariate joint model for multiple
  longitudinal outcomes and a time-to-event.
\newblock {\em Statistics in medicine}, 30(12):1366--1380, 2011.

\bibitem{rizopoulos2014combining}
Dimitris Rizopoulos, Laura~A Hatfield, Bradley~P Carlin, and Johanna~JM
  Takkenberg.
\newblock Combining dynamic predictions from joint models for longitudinal and
  time-to-event data using bayesian model averaging.
\newblock {\em Journal of the American Statistical Association},
  109(508):1385--1397, 2014.

\bibitem{rizopoulos2022jmbayes2}
Dimitris Rizopoulos, Grigorios Papageorgiou, and P~Miranda~Afonso.
\newblock Jmbayes2: extended joint models for longitudinal and time-to-event
  data.
\newblock {\em R package version 0.3--0, ed}, 2022.

\bibitem{rizopoulos2020package}
Dimitris Rizopoulos, Maintainer~Dimitris Rizopoulos, MASS Imports, JAGS
  SystemRequirements, and LinkingTo Rcpp.
\newblock Package ‘jmbayes’.
\newblock {\em Journal of Statistical Software CRAN}, 2020.

\bibitem{rizopoulos2016personalized}
Dimitris Rizopoulos, Jeremy~MG Taylor, Joost Van~Rosmalen, Ewout~W Steyerberg,
  and Johanna~JM Takkenberg.
\newblock Personalized screening intervals for biomarkers using joint models
  for longitudinal and survival data.
\newblock {\em Biostatistics}, 17(1):149--164, 2016.

\bibitem{rosner2021bayesian}
Gary~L Rosner, Purushottam~W Laud, and Wesley~O Johnson.
\newblock {\em Bayesian thinking in biostatistics}.
\newblock CRC Press, 2021.

\bibitem{ross2012simulation}
Sheldon~M Ross.
\newblock Simulation, 5th edision, 2012.

\bibitem{rustand2022denisrustand}
Denis Rustand, Elias~T Krainski, and Haavard Rue.
\newblock Denisrustand/inlajoint: Joint modeling multivariate longitudinal and
  time-to-event outcomes with inla.
\newblock {\em Github}, 2022.

\bibitem{sinha1993semiparametric}
Debajyoti Sinha.
\newblock Semiparametric bayesian analysis of multiple event time data.
\newblock {\em Journal of the American Statistical Association},
  88(423):979--983, 1993.

\bibitem{smith2007boa}
Brian~J Smith.
\newblock boa: an r package for mcmc output convergence assessment and
  posterior inference.
\newblock {\em Journal of statistical software}, 21:1--37, 2007.

\bibitem{speckman2003fully}
Paul~L Speckman and Dongchu Sun.
\newblock Fully bayesian spline smoothing and intrinsic autoregressive priors.
\newblock {\em Biometrika}, 90(2):289--302, 2003.

\bibitem{spiegelhalter2007openbugs}
David Spiegelhalter, Andrew Thomas, Nicky Best, and Dave Lunn.
\newblock Openbugs user manual.
\newblock {\em Version}, 3(2):2007, 2007.

\bibitem{sturtz2019r2openbugs}
Sibylle Sturtz, Uwe Ligges, and Andrew Gelman.
\newblock R2openbugs: a package for running openbugs from r.
\newblock {\em R Package Version}, pages 3--2, 2019.

\bibitem{su2015package}
Yu-Sung Su, Masanao Yajima, Maintainer Yu-Sung Su, and JAGS SystemRequirements.
\newblock Package ‘r2jags’.
\newblock {\em R package version 0.03--08, URL http://CRAN. R-project.
  org/package= R2jags}, 2015.

\bibitem{sylvestre2008comparison}
Marie-Pierre Sylvestre and Michal Abrahamowicz.
\newblock Comparison of algorithms to generate event times conditional on
  time-dependent covariates.
\newblock {\em Statistics in Medicine}, 27(14):2618--2634, 2008.

\bibitem{sylvestre2017permalgo}
MP~Sylvestre, T~Evans, T~MacKenzie, and M~Abrahamowicz.
\newblock Permalgo: Permutational algorithm to generate event times conditional
  on a covariate matrix including time-dependent covariates.
\newblock {\em R package version 1.1, URL https://CRAN.R-project.org/
  package=PermAlgo.}, 2017.

\bibitem{tanton2005encyclopedia}
James Tanton.
\newblock {\em Encyclopedia of mathematics}.
\newblock Facts On File, Inc, 2005.

\bibitem{taylor2013real}
Jeremy~MG Taylor, Yongseok Park, Donna~P Ankerst, Cecile Proust-Lima, Scott
  Williams, Larry Kestin, Kyoungwha Bae, Tom Pickles, and Howard Sandler.
\newblock Real-time individual predictions of prostate cancer recurrence using
  joint models.
\newblock {\em Biometrics}, 69(1):206--213, 2013.

\bibitem{tsodikov_ptc_1998}
Alexander Tsodikov.
\newblock A {Proportional} {Hazards} {Model} {Taking} {Account} of
  {Long}-{Term} {Survivors}.
\newblock {\em Biometrics}, 54(4):1508--1516, 1998.

\bibitem{wang2018bayesian}
Xiaofeng Wang, Yu~Ryan Yue, and Julian~J Faraway.
\newblock {\em Bayesian regression modeling with INLA}.
\newblock CRC Press, 2018.

\bibitem{williamson2008joint}
Paula~R Williamson, Ruwanthi Kolamunnage-Dona, Pete Philipson, and Anthony~G
  Marson.
\newblock Joint modelling of longitudinal and competing risks data.
\newblock {\em Statistics in medicine}, 27(30):6426--6438, 2008.

\bibitem{xu2020semi}
Cong Xu, Pantelis~Z Hadjipantelis, and Jane-Ling Wang.
\newblock Semi-parametric joint modeling of survival and longitudinal data: the
  r package jsm.
\newblock {\em Journal of Statistical Software}, 93:1--29, 2020.

\bibitem{yu_joint_2004}
Menggang Yu, Ngayee~J. Law, Jeremy M.~G. Taylor, and Howard~M. Sandler.
\newblock Joint longitudinal-survival-cure models and their application to
  prostate cancer.
\newblock {\em Statistica Sinica}, 14(3):835--862, 2004.

\bibitem{yu_individual_2008}
Menggang Yu, Jeremy M.~G. Taylor, and Howard~M. Sandler.
\newblock Individual {Prediction} in {Prostate} {Cancer} {Studies} {Using} a
  {Joint} {Longitudinal} {Survival}-{Cure} {Model}.
\newblock {\em Journal of the American Statistical Association},
  103(481):178--187, 2008.

\bibitem{zhang2022joint}
Ningshan Zhang and Jeffrey~S Simonoff.
\newblock Joint latent class trees: A tree-based approach to modeling
  time-to-event and longitudinal data.
\newblock {\em Statistical Methods in Medical Research}, 31(4):719--752, 2022.

\bibitem{zhou2015longitudinal}
Yanjiao Zhou, Gururaj Shan, Erica Sodergren, George Weinstock, W~Allan Walker,
  and Katherine~E Gregory.
\newblock Longitudinal analysis of the premature infant intestinal microbiome
  prior to necrotizing enterocolitis: a case-control study.
\newblock {\em PloS one}, 10(3):e0118632, 2015.

\end{thebibliography}

\newpage
\section*{Supplementary Material A: Gaussian quadrature }
\setcounter{table}{0}
\renewcommand{\thetable}{A. \arabic{table}}
\setcounter{figure}{0}
\renewcommand{\thefigure}{A. \arabic{figure}}
\setcounter{lstlisting}{0}
\renewcommand{\lstlistingname}{Listing A.}
In numerical analysis, Gauss–Legendre quadrature  \cite{bogaert2014iteration} and Gauss–Kronrod quadrature  \cite{tanton2005encyclopedia} are two forms of Gaussian quadrature that can be  used to approximate the definite integral of a function ($ \int _{a}^{b}f(x)\,dx$). These types of integrals can be approximated using $K$-point Gaussian quadrature as follows:
$$\displaystyle \int _{a}^{b}f(x)\,dx\approx \sum _{k=1}^{K}w_{k}f(x_{k}),$$
where $w_i$s and $x_i$s are the weights and points/nodes at which the function $f(x)$ should be evaluated.
Although there are different numbers of knots available, this paper considers $K=15$ nodes. For Gauss-Legendre quadrature with 15 nodes in the interval [-1,1], the points defined for integration can be obtained by the R code given in listing A. 1.
\begin{lstlisting}[caption=The R code  for the Gauss–Legendre quadrature's nodes and weights.,label=Legendre]
# Gauss-Legendre quadrature (15 points)
library(statmod)
glq <- gauss.quad(15, kind = "legendre")
xk <- glq$nodes # Nodes
wk <- glq$weights # Weights
K <- length(xk) # K-points
\end{lstlisting}
For integrating over a general real interval ${\displaystyle [a,b]}$, a change of interval can be applied to convert the problem to one of integrating over
${\displaystyle [-1,1]}$ as $${\displaystyle x_{k,{\text{scaled}}}={\frac {x_{k}+1}{2}}(b-a)+a},$$ and
$${\displaystyle w_{k,{\text{scaled}}}=w_{k}{\frac {b-a}{2}}}.$$
In the package \texttt{JM}  \cite{JM}, a Gauss-Kronrod quadrature with 15 nodes is used to approximate the integral. To utilize the Gauss-Kronrod quadrature, one can refer to \url{https://github.com/tbaghfalaki/Gauss-Kronrod-weights-and-nodes/blob/main/G_Kronrod_15.R}.

\end{document}